\documentclass[useAMS]{mn2e} 

\usepackage{latexsym,graphicx}

\usepackage{color}

\newcommand\kms{{\rm\,km\,s^{-1}}}
\newcommand\msun{\rm\,M_\odot}
\newcommand\rsun{\rm\,R_\odot}
\newcommand\lsun{\rm\,L_\odot}

\newcommand\myr{\msun \, {\rm yr}^{-1}}

\def\apgt{\ {\raise-.5ex\hbox{$\buildrel>\over\sim$}}\ }
\def\aplt{\ {\raise-.5ex\hbox{$\buildrel<\over\sim$}}\ }

\title[Wray\,15-906: a candidate LBV]{Wray\,15-906: a candidate luminous blue variable
discovered with {\it WISE}, {\it Herschel} and SALT}
\author[O. V.~Maryeva et al.]
       {O. V.~Maryeva,$^{1,2}$\thanks{E-mail: olga.maryeva@asu.cas.cz (OVM)} 
       V. V.~Gvaramadze,$^{2,3}$\thanks{E-mail: vgvaram@mx.iki.rssi.ru (VVG)} 
       A. Y.~Kniazev$^{4,5,2}$ and L. N.~Berdnikov$^{2}$ \\
       $^{1}$Astronomical Institute, Czech Academy of Sciences, Fri\v{c}ova 298, 251 65 Ond\v{r}ejov, Czech Republic \\
       $^{2}$Sternberg Astronomical Institute, Lomonosov Moscow State University, Universitetskij Pr. 13, 
       Moscow 119992, Russia\\
       $^{3}$E. Kharadze Georgian National Astrophysical Observatory, Abastumani 0301, Georgia \\
       $^{4}$South African Astronomical Observatory, PO Box 9, 7935 Observatory, Cape Town, South Africa \\
       $^{5}$Southern African Large Telescope Foundation, PO Box 9, 7935 Observatory, Cape Town, South Africa \\
               }
\begin{document}

\date{Accepted 2020 August 27. Received 2020 August 27; in original form 2020 May 12}


\maketitle

\label{firstpage}

\begin{abstract}
We present the results of study of the Galactic candidate luminous blue variable Wray\,15-906, 
revealed via detection of its infrared circumstellar shell (of $\approx2$\,pc in diameter) with 
the {\it Wide-field Infrared Survey Explorer} ({\it WISE}) and the {\it Herschel Space Observatory}. 
Using the stellar atmosphere code {\sc cmfgen} and the {\it Gaia} parallax, we found that 
Wray\,15-906 is a relatively low-luminosity, $\log(L/\lsun)\approx5.4$, star of temperature of 
$25\pm2$\,kK, with a mass-loss rate of $\approx3\times10^{-5} \, \myr$, a wind velocity of 
$280\pm50 \, \kms$, and a surface helium abundance of $65\pm2$ per cent (by mass). In the 
framework of single star evolution, the obtained results suggest that Wray\,15-906 is a post-red 
supergiant star with initial mass of $\approx25 \, \msun$ and that before exploding as a 
supernova it could transform for a short time into a WN11h star. Our spectroscopic monitoring  
with the Southern African Large Telescope (SALT) does not reveal significant changes in the 
spectrum of Wray\,15-906 during the last 8 yr, while the $V$-band light curve of this star over 
years 1999--2019 shows quasi-periodic variability with a period of $\approx1700$\,d and an 
amplitude of $\approx0.2$\,mag. We estimated the mass of the shell to be $2.9\pm0.5 \, \msun$ 
assuming the gas-to-dust mass ratio of 200. The presence of such a shell indicates that 
Wray\,15-906 has suffered substantial mass loss in the recent past. We found that the open star 
cluster C1128-631 could be the birth place of Wray\,15-906 provided that this star is a rejuvenated 
product of binary evolution (a blue straggler). 
\end{abstract}

\begin{keywords}
line: identification -- circumstellar matter -- stars: emission-line, Be -- stars: evolution 
-- stars: individual: Wray\,15-906 -- stars: massive
\end{keywords}

\section{Introduction}
\label{sec:intro}

Sky surveys produced by the {\it Spitzer Space Telescope} (Werner et al. 2004), the {\it Wide-field 
Infrared Survey Explorer} ({\it WISE}; Wright et al. 2010) and the {\it Herschel Space Observatory} 
(Pilbratt et al. 2010) provide rich material for studying the interaction of massive and less massive 
stars with their environment. The high angular resolution and sensitivity of the data obtained by 
these space missions allowed ones to detect a huge number of previously unknown circumstellar nebulae 
(Gvaramadze, Kniazev \& Fabrika 2010b; Mizuno et al. 2010; Watcher et al. 2010; Cox et al. 2012;
Hutsem\'ekers, Cox \& Vamvatira-Nakou 2013), bow shocks (Gvaramadze et al. 2011; Peri et al. 2012; 
Cox et al. 2012; Kobulnicky et al. 2016), dust waves (Ochsendorf et al. 2014) and planetary nebulae 
(Kronberger et al. 2016). Follow-up spectroscopy of central stars associated with the newly detected 
nebulae and bow shocks led to the discovery of rare types of massive stars, such as the luminous blue
variable (LBV) and Wolf-Rayet stars (e.g. Gvaramadze et al. 2009, 2010a; 2012; Wachter et al. 2010, 
2011; Mauerhan et al. 2010; Stringfellow et al. 2012a,b; Burgemeister et al. 2013; Flagey et al. 2014; 
Silva et al. 2017; Cochetti et al. 2020), a massive spun-up and rejuvenated high-velocity runaway star 
(Gvaramadze et al. 2019a), a massive white-dwarf merger product (Gvaramadze et al. 2019b), and a post-very 
late thermal pulse star (Gvaramadze et al. 2019c).

Of the massive stars revealed through the detection of their circumstellar nebulae, perhaps 
the most interesting are the LBVs (Conti 1984; Humphreys \& Davidson 1994). The defining characteristics 
of this rare class of stars\footnote{Only 17 bona fide LBVs are known to date in the Milky Way (Kniazev, 
Gvaramadze \& Berdnikov 2015; Richardson \& Mehner 2018).} are irregular changes in their spectral 
appearance and brightness on time-scales from months to tens of years, leading to blue-red-blue excursions 
on the Hertzsprung-Russell diagram and visual brightness variability by more than one mag (e.g. Humphreys 
\& Davidson 1994; Stahl et al. 2001; van Genderen 2001; Weis \& Bomans 2020). Another common characteristic 
of the majority of LBVs is the presence of pc-scale circumstellar nebulae of various shapes around them 
(Nota et al. 1995; Clark, Larionov \& Arkharov 2005). Currently, about 70 per cent of Galactic bona fide 
LBVs are known to be associated with such nebulae (Kniazev et al. 2015). 

However, despite the great interest in these stars it is not yet settled at what stage of the evolution 
of massive stars the LBV activity takes place and what role the duplicity of massive stars may play in 
the origin of this activity (e.g. Langer et al. 1994; Stothers \& Chin 1996; Groh et al. 2014; Justham, 
Podsiadlowski \& Vink 2014). Also, the driving mechanism(s) of the LBV activity, the role of this activity 
in the evolution of massive stars, and the initial masses of stars susceptible to this activity are still 
poorly understood and remain debated (e.g. Smith, Vink \& de Koter 2004; Vink 2012; Smith 2014; Davidson 
2020).

There are also known several dozens of stars with LBV-like spectra, which did not show major changes in 
their spectra and brightness and which are treated as candidate LBVs (cLBVs; Clark et al. 2005; Gvaramadze 
\& Kniazev 2017; Richardson \& Mehner 2018). Most of these stars are surrounded by compact nebulae similar 
to those detected around bona fide LBVs, which suggests that they might be either dormant or ex-LBVs 
(Bohannan 1997). Spectroscopic and photometric monitoring of cLBVs could potentially confirm that some of 
them are indeed bona fide LBVs, while discovery of new (c)LBVs may allow us to advance our understanding of 
the nature of the LBV phenomenon. 

In this paper, we present the results of study of the new Galactic cLBV Wray\,15-906, 
revealed through the detection of its circular circumstellar shell with {\it WISE} and {\it Herschel}. 
In Section\,\ref{sec:neb}, we present images of the shell and its central star at several 
wavelengths, and review the existing data on the star. In Section\,\ref{sec:obs}, we describe our 
spectroscopic observations of Wray\,15-906 with the Southern African Large Telescope (SALT) and present 
its light curve based on archival and our own photometric observations. In Section\,\ref{sec:ana}, we 
derive fundamental parameters of Wray\,15-906 using the stellar atmosphere code {\sc cmfgen}. In 
Section\,\ref{sec:var}, we discuss the spectroscopic and photometric variability of this star. 
In Section\,\ref{sec:dis}, we discuss the evolutionary status and possible fate of Wray\,15-906, 
estimate the mass of its circumstellar shell, and search for its possible birth cluster. We summarize 
in Section\,\ref{sec:sum}.

\begin{figure*}
\begin{center}
\includegraphics[width=17cm,angle=0]{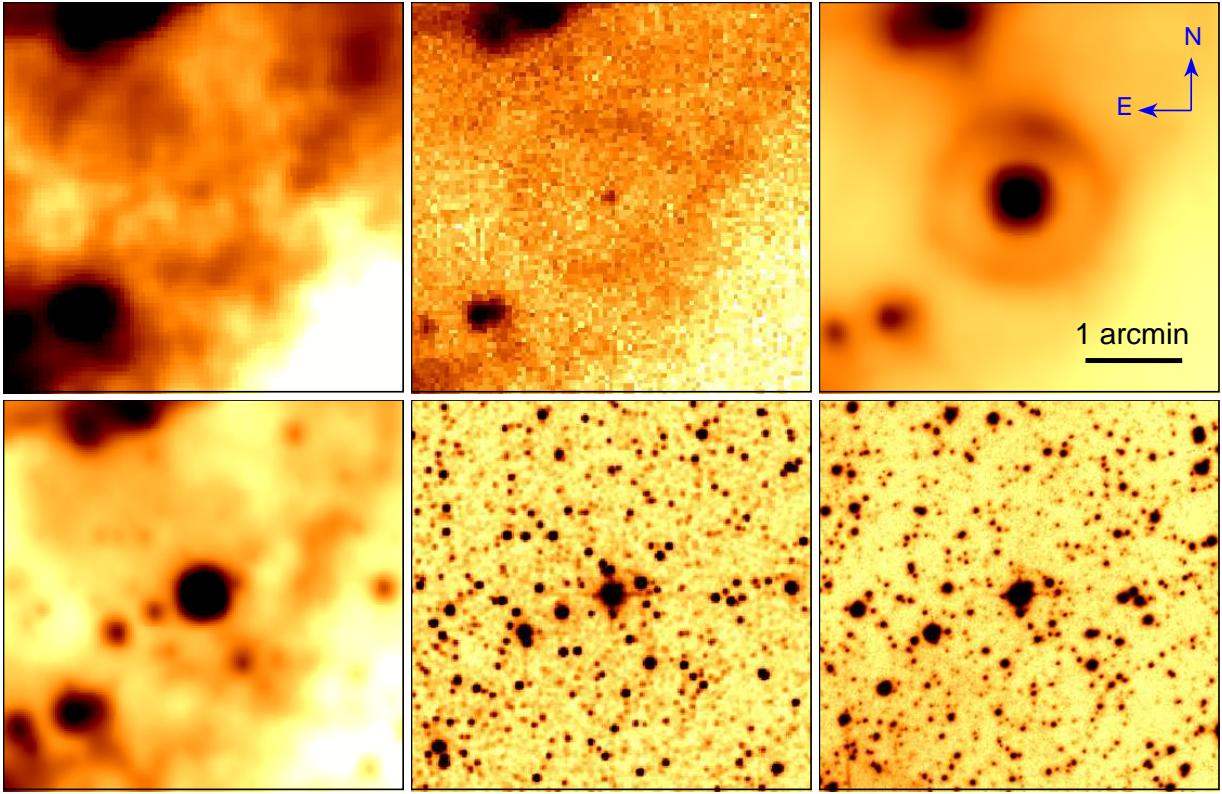}
\end{center}
\caption{From left to right, and from top to bottom: {\it Herschel} PACS 160 and 70 \micron,
{\it WISE} 22 and 12 \micron, 2MASS $K_{\rm s}$-band and SHS H\,$\alpha$+[N\,{\sc ii}] images 
of the region containing Wray\,15-906 and the shell around it (the scale and orientation of 
the images are the same). At a distance of 3.53 kpc, 1 arcmin corresponds to $\approx1$ pc.}
\label{fig:neb}
\end{figure*}

\section{Wray\,15-906 and its circular circumstellar shell}
\label{sec:neb}

Wray\,15-906 (also ALS\,2533 and Hen\,3-729) was identified as an emission-line star by Wray (1966).
Later on, it was classified as a possible Wolf-Rayet star by Stephenson \& Sanduleak (1971) and 
Stenholm (1975), and as a possible symbiotic star by Henize (1976). Although Wray\,15-906 is indicated 
in the SIMBAD data base\footnote{http://simbad.u-strasbg.fr/simbad/} as a possible Wolf-Rayet star, it 
remained unexplored until recently and its nature was unclear.

Wray\,15-906 attracted our attention after we discovered a circular shell around it using {\it WISE}
data (Kniazev \& Gvaramadze 2015). Follow-up optical spectroscopy of Wray\,15-906 with SALT in 2012
(Kniazev \& Gvaramadze 2015) revealed a rich emission spectrum typical of hot LBV stars, which along 
with the presence of the circular circumstellar shell point to the possibility that this star is an
LBV. Subsequent spectroscopic monitoring of Wray\,15-906 (carried out in the period 2013--2016) and the 
available photometric data (covering the time period from 2001 to 2016), however, did not reveal 
significant changes in the spectrum and brightness of this star, which led us to consider it as a 
cLBV (Kniazev, Gvaramadze \& Berdnikov 2017).

In Fig.\,\ref{fig:neb}, we show {\it Herschel} 160 and 70 \micron, {\it WISE} 22 and 12\,\micron, 
Two-Micron All Sky Survey (2MASS; Skrutskie et al. 2006) $K_{\rm s}$-band, and SuperCOSMOS H-alpha 
Survey (SHS; Parker et al. 2005) H\,$\alpha$+[N\,{\sc ii}] images of the region containing 
Wray\,15-906 and the nebula around it. In the discovery {\it WISE} 22\,\micron \, 
image\footnote{Shown for the first time in Kniazev \& Gvaramadze (2015).} (see Fig.\,\ref{fig:neb}), 
the nebula appears as a circular limb-brightened shell of angular radius of $\approx1$ arcmin, which 
at a distance of $d=3.53$\,kpc (see below) corresponds to $\approx1$\,pc (which is typical of 
circumstellar shells around LBVs; e.g. Weis 2001). There is no obvious counterpart to the shell at 
shorter infrared wavelengths, but it is clearly visible in the 70\,\micron \, image obtained with the 
Photodetector Array Camera and Spectrometer (PACS) instrument on board the {\it Herschel Space 
Observatory} (Pilbratt et al. 2010). PACS 160\,\micron \, and {\it WISE} 12\,\micron \, images show that 
Wray\,15-906 is surrounded by clumpy emission on all sides, but it is not clear whether it is physically 
associated with the star. Wray\,15-906 was also covered by the Galactic Legacy Infrared Mid-Plane Survey 
Extraordinaire (GLIMPSE; Benjamin et al. 2003), carried out by the {\it Spitzer} Infrared Array Camera 
(IRAC; Fazio et al. 2004). The shell is not visible in none of the four (8, 5.8, 4.5 and 3.6\,\micron) 
images provided by this survey. It is also not visible in the SHS H\,$\alpha$+[N\,{\sc ii}] image.

In Table\,\ref{tab:det} we summarize basic properties of Wray\,15-906, which we will later use for 
spectral modelling and searching for possible birth cluster of the star. The coordinates, 
parallax and proper motion measurements are from the second {\it Gaia} Data Release (DR2; Gaia 
Collaboration et al. 2018), while the distance is derived from a simple inversion of the parallax. 
The $B$, $V$ and $I_{\rm c}$ magnitudes were measured by us in the course of photometric monitoring 
of central stars of the infrared nebulae discovered with {\it Spitzer} and {\it WISE} (see 
Section\,\ref{sec:pho}). They were obtained closest in time (about one and half month earlier) to 
the spectrum of Wray\,15-906 used in Section\,\ref{sec:mod} for spectral modelling. The $J$, $H$ and 
$K_{\rm s}$ magnitudes are from the 2MASS All-Sky Catalog of Point 
Sources (Cutri et al. 2003). The {\it WISE} (3.4, 4.6, 12 and 22\,\micron) photometry is from the 
AllWISE Data Release (Cutri et al. 2014). The IRAC magnitudes at 3.6, 4.5, 5.8 and 8\,\micron \, are 
from the GLIMPSE Source Catalog (I+II+3D) (Spitzer Science Center 2009). Note the big difference 
between the {\it WISE} 3.4\,\micron \, and IRAC 3.6\,\micron \, magnitudes. Since it is unlikely that 
this difference is caused by the stellar variability (see Section\,\ref{sec:pho}) and because the 
{\it WISE} 3.4\,\micron \, magnitude better fits the spectral energy distribution (SED) of 
Wray\,15-906 (see Section\,\ref{sec:mod} and Fig.\,\ref{fig:sed}), we treat the IRAC 3.6\,\micron 
\, magnitude as unreliable.

Note also that the distance error bar given in Table\,\ref{tab:det} corresponds to the 
formal 1$\sigma$ uncertainty of the parallax. In fact, the range of allowed distances to Wray\,15-906
should be much broader because of the large astrometric excess noise of this star (e.g. Lindegren et al. 
2018; Smith et al. 2019), which is only about 1.6 times less than the parallax and about 3.6 times larger 
than the parallax uncertainty (Gaia Collaboration et al. 2018). Thus, the distance given in 
Table\,\ref{tab:det} should be considered as a crude estimate, which hopefully will be better constrained 
by the forthcoming third {\it Gaia} data release.

\begin{table}
  \centering{\caption{Properties of Wray\,15-906.}
  \label{tab:det}
 \begin{tabular}{lcc}
    \hline
  RA(J2000) & $11^{\rm h} 54^{\rm m} 43\fs54$ & {\it Gaia} DR2 \\ 	
  Dec.(J2000) & $-63\degr 13\arcmin 31\farcs1$ & {\it Gaia} DR2 \\ 
  $l$ & $296\fdg5959$ & {\it Gaia} DR2 \\
  $b$ & $-1\fdg0534$ & {\it Gaia} DR2 \\
  $\pi$ (mas) & $0.2835\pm0.0495$ & {\it Gaia} DR2 \\
  $\mu_\alpha$ (mas) & $-5.820\pm0.090$ & {\it Gaia} DR2 \\
  $\mu_\delta$ (mas) & $1.158\pm0.071$ & {\it Gaia} DR2 \\
  $d$ (kpc) & $3.53^{+0.74} _{-0.53}$ & {\it Gaia} DR2 \\
  $B$ (mag) & $14.34\pm0.03$ & This paper \\
  $V$ (mag) & $12.50\pm0.03$ & This paper  \\
  $I_{\rm c}$ (mag) & $9.86\pm0.02$ & This paper \\
  $J$ (mag) & $7.70\pm0.02$ & 2MASS \\  
  $H$ (mag) & $6.87\pm0.03$ & 2MASS \\
  $K_{\rm s}$ (mag) & $6.24\pm0.03$ & 2MASS \\
  3.4 \micron \, (mag) & $5.35\pm0.18$ & {\it WISE} \\	
  3.6 \micron \, (mag) & $7.26\pm0.32$ & {\it Spitzer} IRAC \\ 
  4.5 \micron \, (mag) & $4.99\pm0.05$ & {\it Spitzer} IRAC \\
  4.6 \micron \, (mag) & $4.69\pm0.09$ & {\it WISE} \\	
  5.8 \micron \, (mag) & $4.60\pm0.03$ & {\it Spitzer} IRAC \\
  8 \micron \, (mag)   & $4.18\pm0.03$ & {\it Spitzer} IRAC \\ 
  12 \micron \, (mag)  & $3.99\pm0.02$ & {\it WISE} \\	
  22 \micron \, (mag)  & $2.97\pm0.03$ & {\it WISE} \\
  \hline
 \end{tabular}
}
\end{table}

\begin{figure*}
\begin{center}
\includegraphics[width=12.5cm,angle=270]{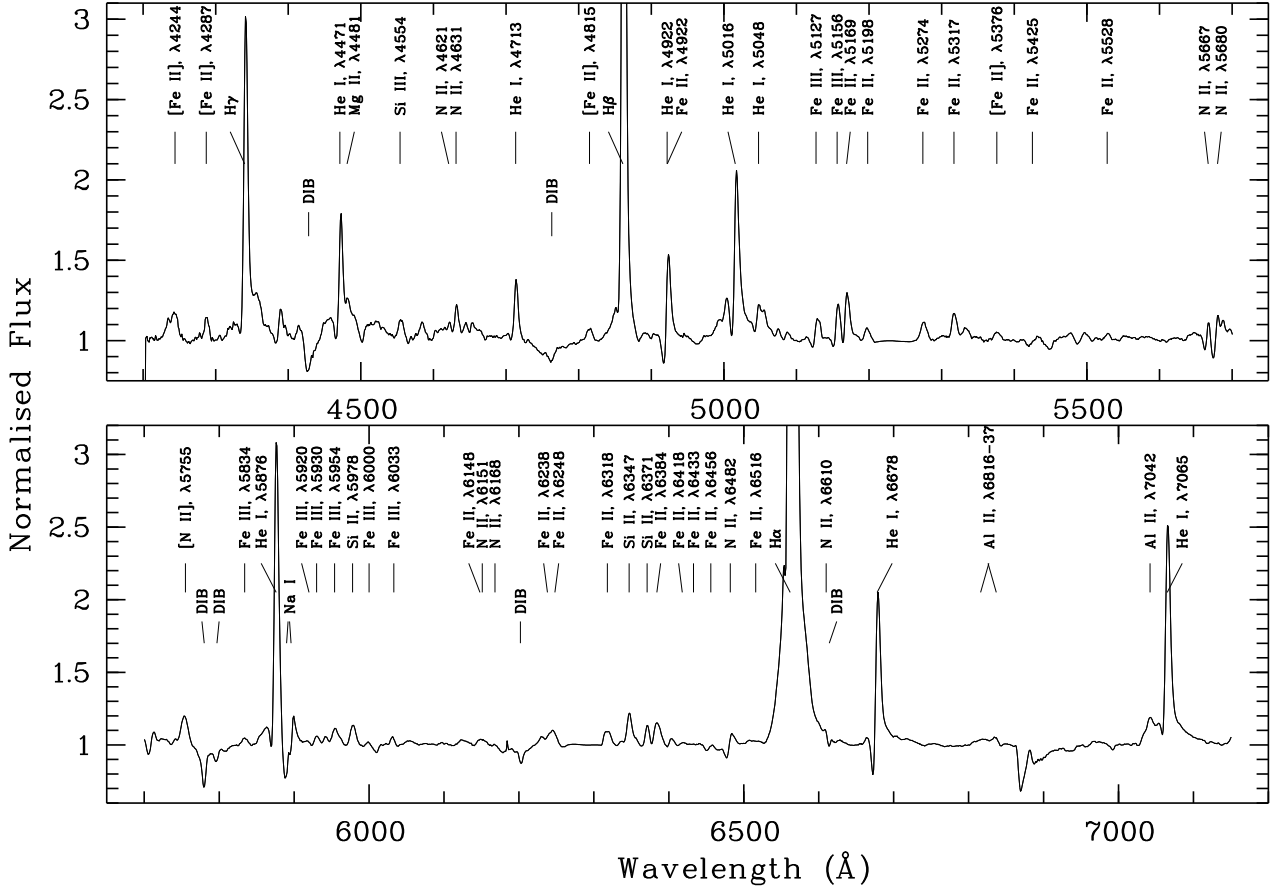}
\end{center}
\caption{RSS spectrum of Wray\,15-906 obtained on 2012 February 19, with the major lines 
and diffuse interstellar bands (DIBs) identified.}
\label{fig:spe}
\end{figure*}

\section{Observations}
\label{sec:obs}

\subsection{Spectroscopy}
\label{sec:spe}

\begin{table*}
\caption{Journal of the SALT RSS and HRS observations of Wray\,15-906.}
\label{tab:obs}
\begin{tabular}{llccccccc} \hline
Spectrograph & Date & Exposure & Spectral scale & Resolving & PA & Seeing & Spectral range \\
 & & (sec) & (\AA \, pixel$^{-1}$) & power & ($\degr$) & (arcsec) & (\AA) \\
 \hline
RSS & 2012 February 19 & 60+300 & 0.97  & 1000 & 0   &  1.2 & 4200$-$7300 \\
RSS & 2013 April 22    & 10+300 & 0.97  & 1000 & 42  &  1.6 & 4200$-$7300 \\
RSS & 2013 May 25      & 60+300 & 0.97  & 1000 & 42  &  2.4 & 4200$-$7300 \\
RSS & 2013 June 22     & 60+600 & 0.97  & 1000 & 42  &  1.7 & 4200$-$7300 \\
RSS & 2014 March 14    & 10+300 & 0.97  & 1000 & 316 &  2.0 & 4200$-$7300 \\
RSS & 2015 May 27      & 10+300 & 0.97  & 1000 & 136 &  2.4 & 4200$-$7300 \\
HRS & 2015 June 13     & 1200   & 0.04  & 16\,000 & --  &  3.0 & 3700$-$8900 \\
HRS & 2016 May 5       & 1200   & 0.04  & 16\,000 & --  &  3.0 & 3700$-$8900 \\
HRS & 2017 April 16    & 1200   & 0.04  & 16\,000 & --  &  3.0 & 3700$-$8900 \\
HRS & 2019 February 18 & 1200   & 0.04  & 16\,000 & --  &  3.0 & 3700$-$8900 \\  
HRS & 2020 February 4  & 2400   & 0.04  & 16\,000 & --  &  3.0 & 3700$-$8900 \\
\hline
\end{tabular}
\end{table*}

We obtained our first spectrum of Wray 15-906 (Fig.\,\ref{fig:spe}) on 2012 February 19 on SALT 
(Buckley, Swart \& Meiring 2006; O'Donoghue et al. 2006). The observation was carried out with 
the Robert Stobie Spectrograph (RSS; Burgh et al. 2003; Kobulnicky et al. 2003) in the long-slit mode.
As expected, the spectrum turned out to be interesting (namely, similar to that of the bona-fide LBV 
P\,Cygni), so that we continued our observations in the following years in the hope of finding 
significant changes in it. In 2013--2015 we obtained five more RSS spectra of Wray\,15-906.

All RSS spectra were taken with a slit width of 1.25 arcsec and with the PG900 grating, which covers the
spectral range of 4200$-$7300~\AA. This choice of the slit width and grating provides a final reciprocal 
dispersion of $0.97$~\AA \, pixel$^{-1}$ and spectral resolution FWHM of 4.4--4.7~\AA \, ($R\approx1000$). 
The observations were done with a short (10 or 60~s) and long (300 or 600~s) exposures to exclude possible 
saturation of strong emission lines. An Xe lamp arc spectrum (taken immediately after the science frames)
was used for wavelength calibration. The relative flux calibration was performed using observations of 
spectrophotometric standard stars. A comparison of the obtained spectra revealed only slight changes in 
their appearance. 

In 2015 the High Resolution Spectrograph (HRS; Barnes et al. 2008; Bramall et al. 2010, 2012; 
Crause et al. 2014) was commissioned at the SALT, which allowed us to continue monitoring of Wray\,15-906 
with a much better spectral resolution. We used this fibre-fed, high dispersion \'echelle spectrograph 
in the low resolution mode ($R\approx$16\,000) to obtain spectra in the blue and red arms over the total 
spectral range of $\approx$3700--8900~\AA. The CCDs of both arms were read out by a single amplifier 
with a 1$\times$1 binning. Four spectra of Wray\,15-906 were taken in 2015--2019 with the same exposure 
time of 1200~s, and one more spectrum on 2020 February 4 with the exposure time of 2400~s. Three arc spectra 
of ThAr lamp and three spectral flats were obtained for each observation under the same resolution mode 
during a weekly set of HRS calibrations. The spectrum obtained in 2016 was used in Section\,\ref{sec:mod}
for spectral modelling and its portions are shown in Figs\,\ref{fig:com} and \ref{fig:mod}. 

The log of our spectroscopic observations is given in Table\,\ref{tab:obs}.

The primary reduction of the RSS and HRS spectra was performed using the SALT science pipeline 
(Crawford et al. 2010). For further reduction of the long-slit data we followed the procedure 
described in Kniazev et al. (2008). The HRS spectra were further reduced using the {\sc midas} 
HRS pipeline described in detail in Kniazev, Gvaramadze \& Berdnikov (2016) and Kniazev et al. 
(2019).  

\begin{figure*}
\begin{center}
\includegraphics[width=9cm,angle=270,clip=]{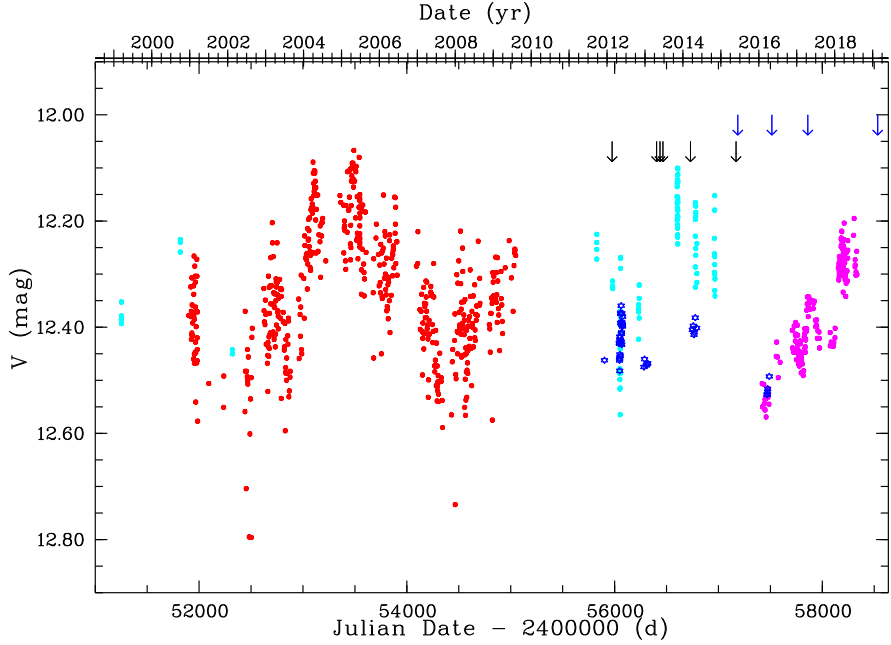}
\end{center}
\caption{$V$-band light curve of Wray\,15-906 as a function of time over years 1999--2019. Red, cyan
and magenta dots correspond, respectively, to the ASAS, OMC and ASAS-SN photometry. Our measurements 
are shown with open blue hexagons. The arrows mark the dates of the RSS (bottom row) and HRS (upper 
row) spectra obtained in 2012--2019.}
\label{fig:pho}
\end{figure*}

\subsection{Photometry}
\label{sec:pho}

As a part of our photometric observations of central stars of the infrared nebulae detected with
{\it Spitzer} and {\it WISE}, we determined $B, V$ and $I_{\rm c}$ magnitudes of Wray\,15-906 
on CCD frames obtained with the South African Astronomical Observatory (SAAO) 76-cm and 1-m 
telescopes in 2011--2016. We used an SBIG ST-10XME camera equipped with filters of the Kron-Cousins 
system (see Berdnikov et al. 2012). 

To construct the long-term light curve of Wray\,15-906, we also collected $V$-band photometry of 
this star from the All Sky Automated Survey (ASAS; Pojmanski 1997) and the All-Sky Automated Survey 
for Supernovae (ASAS-SN; Shappee et al. 2014; Kochanek et al. 2017), as well as that obtained by the 
Optical Monitoring Camera (OMC; Alfonso-Garz{\'o}n et al. 2012) on board the International 
Gamma-Ray Astrophysics Laboratory {\it INTEGRAL} observatory. 

From the ASAS survey we used only `grade A' data with accuracy of $\approx0.04-0.07$ mag. From 
the ASAS-SN and OMC photometry we selected the data points with accuracy better than 0.02 and 0.1 mag, 
respectively. Altogether, we collected 857 measurements covering the years 1999--2019. The resulting 
$V$-band light curve for Wray\,15-906 is presented in Fig.\,\ref{fig:pho}. One can see that 
Wray\,15-906 experiences a quasi-periodic brightness variability of several tenths of mag
around the average value of $V=12.35\pm0.11$ mag (see also Section\,\ref{sec:phot}).

\section{Spectral analysis and stellar parameters}
\label{sec:ana}

\subsection{The stellar spectrum}
\label{sec:ste}

\begin{figure*}
\begin{center}
\includegraphics[width=15cm,angle=0,clip=]{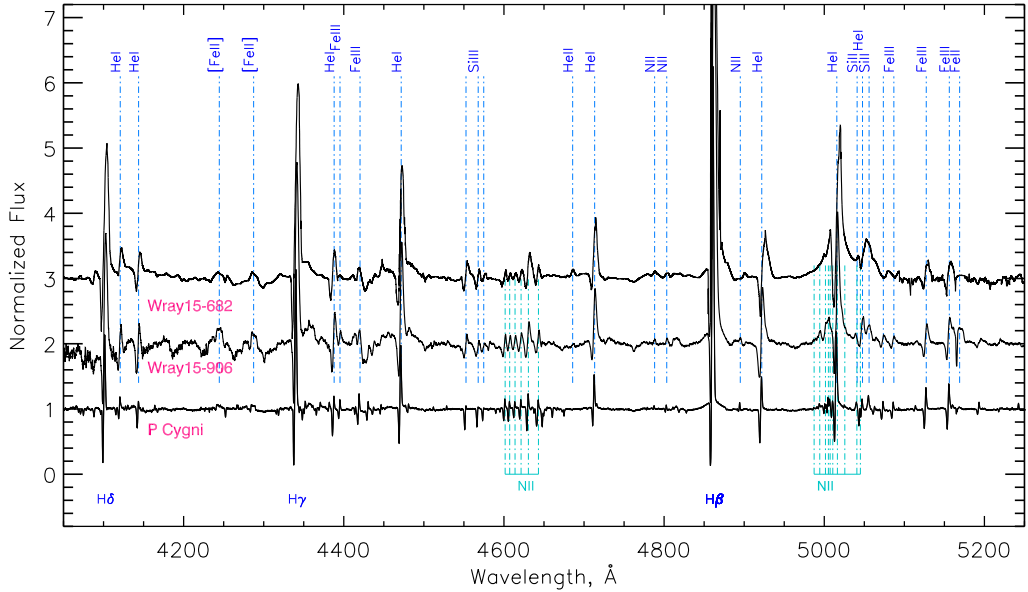}
\end{center}
\caption{Comparison of a normalized portion of the HRS spectrum of Wray\,15-906 taken on 2016 May 5 
with corresponding parts of spectra of the Galactic bona fide LBV P\,Cygni and the WN11h star 
Wray\,15-682.}
\label{fig:com}
\end{figure*}

Wray\,15-906 displays a rich emission-line spectrum, dominated by strong Balmer 
and He\,{\sc i} lines, all of which have P\,Cygni profiles (see Figs\,\ref{fig:com} and
\ref{fig:mod}). No He\,{\sc ii} lines are detected neither in the RSS, nor in the HRS
spectra. Other detected emission lines are due to N\,{\sc ii} (all show P\,Cygni 
profiles), Fe\,{\sc ii} and {\sc iii}, Si\,{\sc ii} and {\sc iii}, Al\,{\sc ii} and Mg\,{\sc ii}. 
Like in P\,Cygni and several other (c)LBVs (Stahl et al. 1993; Gvaramadze et al. 2010a; 2012a), 
the Fe\,{\sc iii} lines with low multiplet numbers show clear P\,Cygni profiles, while those 
with higher multiplet numbers are fully in emission. The near-infrared part of the HRS spectra 
shows Paschen lines in emission and emission lines of neutral nitrogen (N\,{\sc i} $\lambda$8682) 
and oxygen (O\,{\sc i} 7772-5, 8446). The spectra also show the prominent forbidden line of 
[N\,{\sc ii}] $\lambda$5755 and numerous forbidden lines of singly-ionized iron. In the HRS 
spectra these lines have a flat-topped appearance (see Figs\,\ref{fig:mod} and \ref{fig:var}). 

In Fig.\,\ref{fig:com}, we show the blue region of the HRS spectrum of Wray\,15-906 obtained on 
2016 May 5 and compare it with spectra of the bona-fide LBV P\,Cygni and the WN11h (Smith, 
Crowther \& Prinja 1994; Walborn \& Fitzpatrick 2000) star Wray\,15-682 (Hen\,3-519, WR31a)
\footnote{The spectrum of P\,Cygni is taken from the ELODIE archive (Moultaka et al. 2004) at 
Observatoire de Haute-Provence, while that of Wray\,15-682 is from Smith et al. (1994).}. One 
can see that the three spectra bear a strong resemblance with each over. The main 
difference is that Wray\,15-906, like P\,Cygni, has no He\,{\sc ii} lines, meaning that it is 
a P\,Cygni-type B supergiant. On the other hand, the emission lines in Wray\,15-906 and Wray\,15-682 
are much broader than in P\,Cygni, indicating that the wind velocity in the former two stars is higher
than that in the latter one. The results of our spectral modelling (see the next section) show that 
Wray\,15-906 is a blueward evolving post-red supergiant star and that its temperature ($25\pm2$\,kK) 
is a bit lower than that of Wray\,15-682 ($27\pm1$\,kK; Smith et al. 1994). This implies that 
Wray\,15-906 is on the more advanced evolutionary stage than P\,Cygni and that soon it could turn 
into a WN11h star and explode as a supernova (see Section\,\ref{sec:evo}). 

\subsection{Spectral modelling}
\label{sec:mod}

\begin{figure*}
\begin{center}
\includegraphics[width=16cm,angle=90,angle=270,clip=]{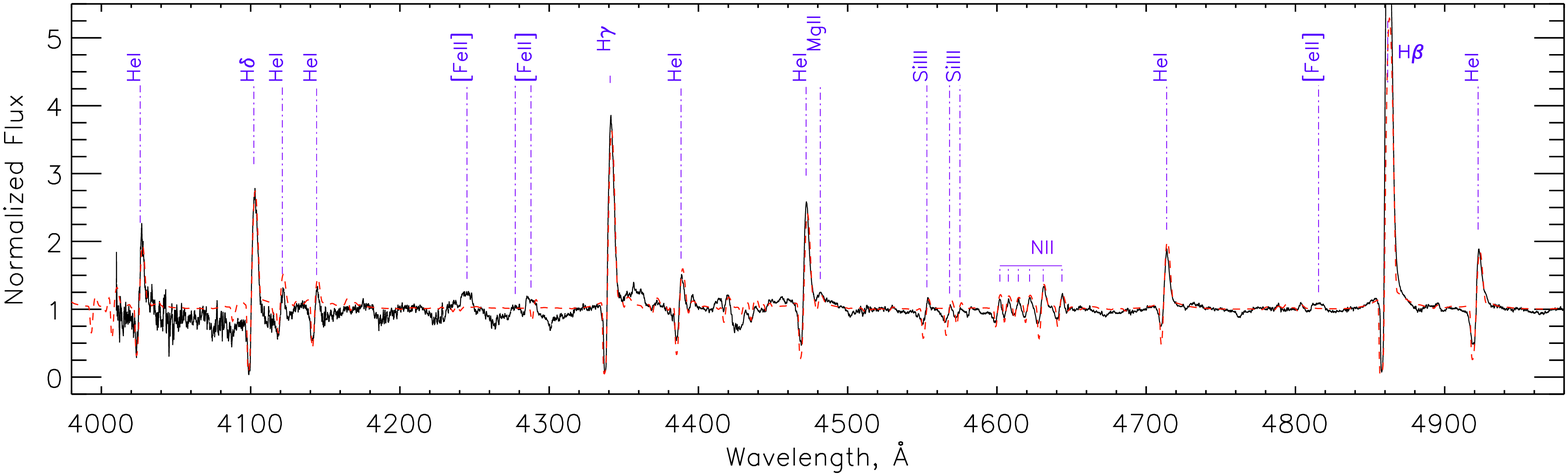}
\includegraphics[width=16cm,angle=90,angle=270,clip=]{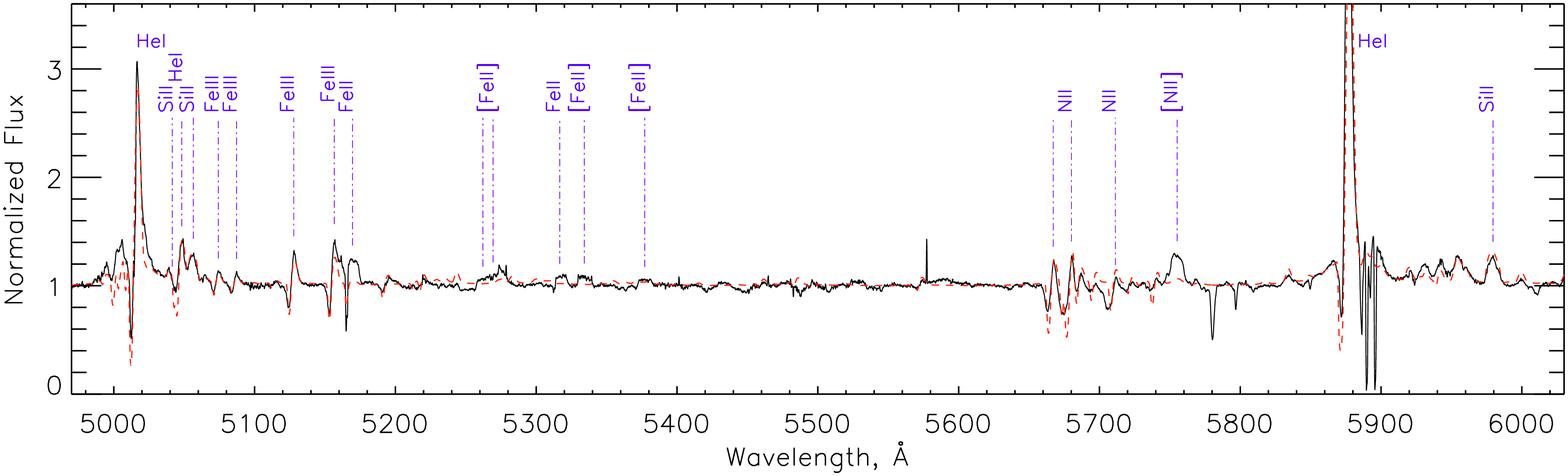}
\includegraphics[width=16cm,angle=90,angle=270,clip=]{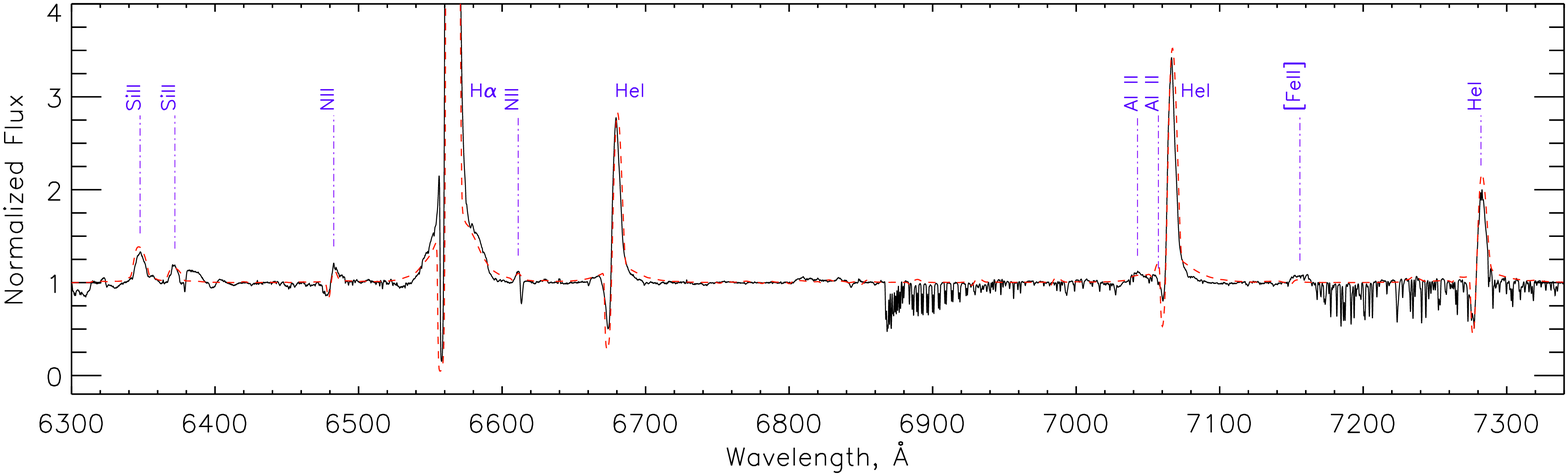}
\includegraphics[width=16cm,angle=90,angle=270,clip=]{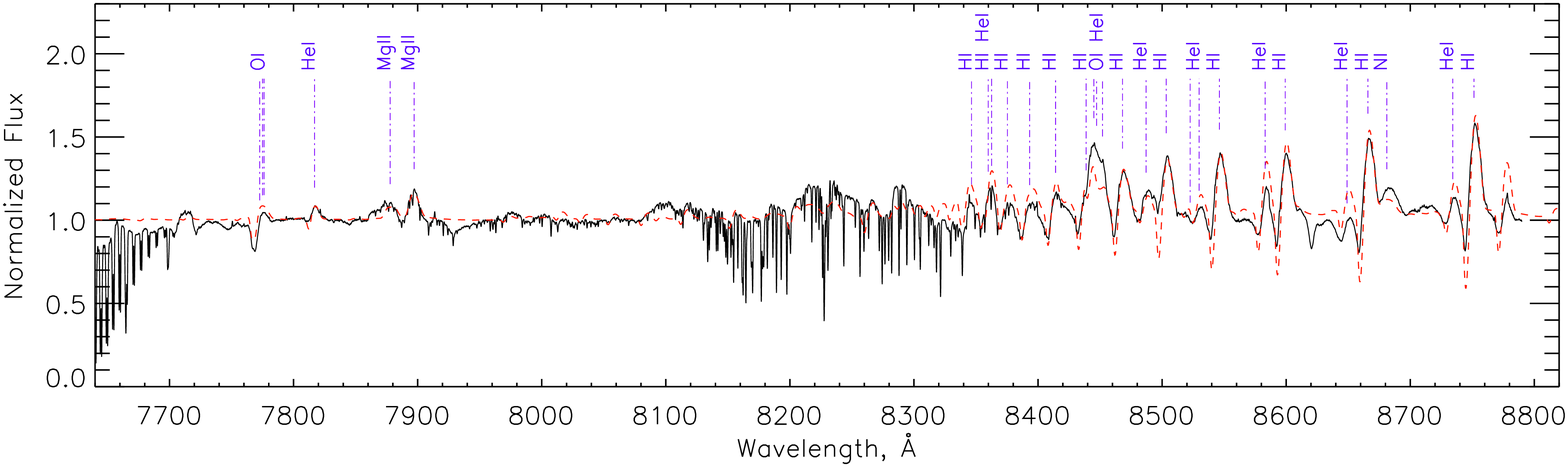}
\end{center}
\caption{Normalized HRS spectrum of Wray\,15-906 taken on 2016 May 5 (black solid line), 
compared with the best-fitting {\sc cmfgen} model (red dashed line) with the parameters 
as given in Table\,\ref{tab:par}.}
\label{fig:mod}
\end{figure*}

The basic parameters of Wray\,15-906 were determined by means of the {\sc cmfgen} code of Hillier 
\& Miller (1998), which was developed to solve radiative transfer equations for objects 
with spherically symmetric extended outflows using the full comoving-frame formulation of the 
radiative transfer equation (e.g. Hillier et al. 2001; Najarro et al. 2009; Groh et al. 2009b). 
{\sc cmfgen} incorporates line blanketing, the effects of clumping and Auger ionization. Every model 
is characterized by a set of parameters, such as the hydrostatic stellar radius $R_*$, luminosity 
$L_*$, mass-loss rate $\dot{M}$, volume filling factor $f$, wind terminal velocity $v_\infty$, 
stellar mass $M_*$, and abundances $Z_i$ of elementary species (our modelling included H, He, C, N, 
O, Ne, Mg, Al, Si, S, Ar, Ca and Fe). Fig.\,\ref{fig:mod} shows the comparison of the best-fitting 
model spectrum with the normalized HRS spectrum obtained on 2016 May 5.
As the length of the HRS \'echelle orders is $\approx140$ \AA, the continuum normalization 
was straightforward even in the region of the H\,$\alpha$ line with its wide ($\approx60$ \AA) wings. 
The spectrum was independently normalized manually by two co-authors using the {\sc midas} and {\sc 
dech30}\footnote{http://gazinur.com/DECH-software.html} software packages. Dividing one normalized 
spectrum by another showed that they are consistent with each other to within 5 per cent. 
This makes us confident that the normalization procedure does not significantly affect the line 
wings and the results of our modelling.

The modelling proceeded in several steps (cf. Gvaramadze et al. 2018). First, we derived the  
temperature $T_*$ at the hydrostatic radius and the temperature $T_{\rm eff}$ at the radius $R_{2/3}$ 
where the Rosseland optical depth is equal to $2/3$. For this, we used the intensities of the 
Si\,{\sc ii}--{\sc iii}, Fe\,{\sc ii}--{\sc iii}, He\,{\sc i} and N\,{\sc i}--{\sc ii} lines. 

\begin{figure*}
\begin{center}
\includegraphics[width=17cm,angle=0]{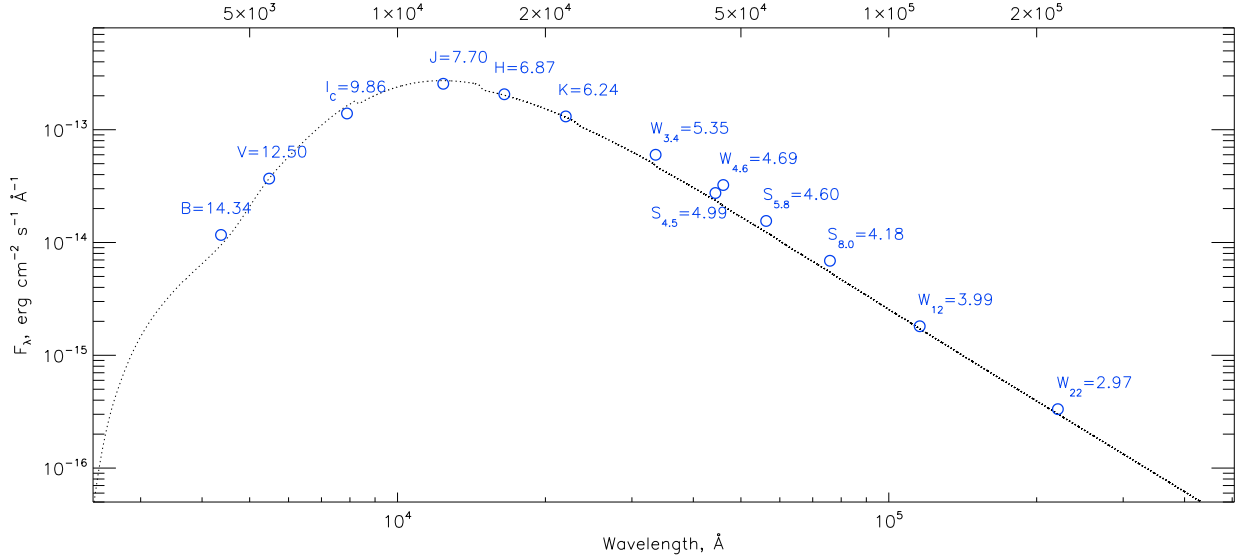}
\end{center}
\caption{Observed flux distribution of Wray\,15-906 (open circles) based on the 
photometric data from Table 1, compared to the continuum of the reddened model
spectrum (dotted line) with the parameters as given in Table\,\ref{tab:par}.}
\label{fig:sed}
\end{figure*}

Then, we derived the colour excess $E(B-V)$ of the star of 2.07 mag. For this, we adjusted the 
observed SED with the model spectrum scaled to the {\it Gaia} DR2 distance to Wray\,15-906 of $d=3.53$ 
kpc (see Fig.~\ref{fig:sed}). We used the extinction law from Fitzpatrick (1999) with a 
total-to-selective absorption ratio of $R_V=3.1$. Note that although $R_V$ was found to vary from 
2.2 to 5.8 between different lines of sight in the Milky Way (Fitzpatrick, 1999), the observed 
SED was fitted pretty well without need to assume a peculiar reddening law.

Next, we determined the radius (and luminosity) of Wray\,15-906. For this, we compared the observed SED 
with the synthetic spectrum calculated for the adopted distance, corrected for interstellar reddening and 
convolved with the transmission curves of the standard Johnson $B, V$ and $I_{\rm c}$ filters. 

On the next step, $v_\infty$ and $\dot{M}$ were estimated by using the lines with P\,Cygni-type profiles and 
intensities of the emission lines. {\sc cmfgen} allows us to build models with different velocity laws 
describing the radial dependence of the wind velocity. For Wray\,15-906 we found it sufficient to use a wind
velocity law exponent of $\beta=1$. 

As a final step, we derived abundances of hydrogen, helium, nitrogen and oxygen. The H and He abundances 
were derived by iterative adjustment of the He-to-H abundance ratio and other fundamental parameters of 
Wray\,15-906 to reproduce the overall shape of all detected lines of these elements. The N abundance was 
derived by analysing the behaviour of the N\,{\sc ii} and N\,{\sc i} lines, while the O\,{\sc i} lines in 
near-infrared part of the spectrum were used to estimate the O abundance. Since no carbon lines were 
detected in the spectrum, for this element we derived only an upper limit on its abundance by decreasing 
its value until the C\,{\sc ii} $\lambda\lambda$7231-36-37 lines disappear in the model spectrum. From the 
large number of Si and Mg lines we estimated the abundances of these elements to be around the solar ones.
For other elements in our models we adopted their solar abundances; in the {\sc cmfgen} code the solar 
abundances are taken from Cox (2000).

The stellar wind was assumed to be clumpy with a void interclump medium (Hillier \& Miller 1999). 
The wind volume filling factor $f(r)=\bar{\rho}/\rho(r)$, where $\bar{\rho}$ is the homogeneous (unclumped) 
wind density and ${\rho}(r)$ is the density in clumps (assumed to be optically thin), depends on radius as 
$f(r)= f_\infty + (1 - f_\infty )\exp(-v(r)/v_{\rm cl})$, where $f_\infty$ characterize the density
contrast and $v_{\rm cl}$ is the velocity at which clumping starts. In our modelling, we adopted 
$f_\infty=0.1$ and $v_{\rm cl}=50 \, \kms$. We also calculated models with higher values of $f_\infty$
of 0.25 and 0.5. However, the increase of the wind homogeneity leads to the increase of the influence of 
electron scattering on line profiles (Hillier 1991) and to the disappearance of absorption components in 
the H\,$\alpha$ and H\,$\beta$ lines (cf. Groh et al 2009b), which is not consistent with the observed 
spectrum. Correspondingly, we used $f_\infty=0.1$ in the final model. Puls et al. (2006) studied the radial 
stratification of clumping in hot star winds by simultaneous modelling of H\,$\alpha$, infrared, mm and 
radio data, and demonstrated that for dense winds it decreases outwards (see also Najarro, Hanson \& Puls 
2011). And indeed our computations show that for simultaneous fitting of the visible and near-infrared parts 
of the spectrum the clumping has to start to disappear at the distance where the wind velocity exceeds 
$200 \, \kms$.

\begin{table}
\caption{Stellar parameters of Wray\,15-906 for three distances (see the text for details).}
\label{tab:par}
\begin{tabular}{lllc}
\hline
$d$ (kpc)                       &   3.00            &    3.53                &     4.27            \\
$\log(L_*/\lsun)$               &   $5.19\pm0.02$   &    $5.38\pm0.02$       &     $5.50\pm0.02$    \\
$\dot{M}$ ($10^{-5} \, \myr$)   &   $2.8\pm0.3$     &    $3.3\pm0.3$         &     $4.0\pm0.3$     \\
$R_*    $ ($\rsun$)             &   $21.0\pm4.4$    &    $26.1\pm5.0$        &     $30.1\pm5.6$    \\
$T_{eff}$ (kK)                  &   $18.7\pm1.5$    &    $19.2\pm1.5$        &     $20.1\pm1.5$    \\
$R_{2/3}$ ($\rsun$)             &   37.5            &    44.3                &     46.3            \\
                                &                   &                        &                     \\
$T_*    $ (kK)                  & \multicolumn{2}{c}{$25\pm 2$}                                    \\                                                
$v_\infty$($\kms$)              & \multicolumn{2}{c}{$280\pm50$ }                   \\
$\beta$ (adopted)               & \multicolumn{2}{c}{1.0 }                          \\
$f$ (adopted)                   & \multicolumn{2}{c}{0.1 }                          \\                                 
H/He                            & \multicolumn{2}{c}{$2.15\pm0.15$}                 \\
H (mass fraction)               & \multicolumn{2}{c}{$0.35\pm0.02$}                 \\
He (mass fraction)              & \multicolumn{2}{c}{$0.65\pm0.02$}                 \\
C (mass fraction)               & \multicolumn{2}{c}{$\leq6\times10^{-5}$ }         \\
N (mass fraction)               & \multicolumn{2}{c}{$(2.7-3.5)\times10^{-3}$}      \\
O (mass fraction)               & \multicolumn{2}{c}{$(0.8-1.6)\times10^{-4}$ }     \\
$E(B-V)$ (mag)                  & \multicolumn{2}{c}{2.07}                          \\
$R_V$ (adopted)                 & \multicolumn{2}{c}{3.1}                           \\
\hline
\end{tabular}
\end{table}

The derived parameters are listed in Table~\ref{tab:par} for thee distances to Wray\,15-906, 
derived from the {\it Gaia} DR2 parallax and its $\pm1$-sigma values (see Table\,\ref{tab:det}).

\begin{figure}
\centering{
 \includegraphics[clip=,angle=0,width=8.5cm]{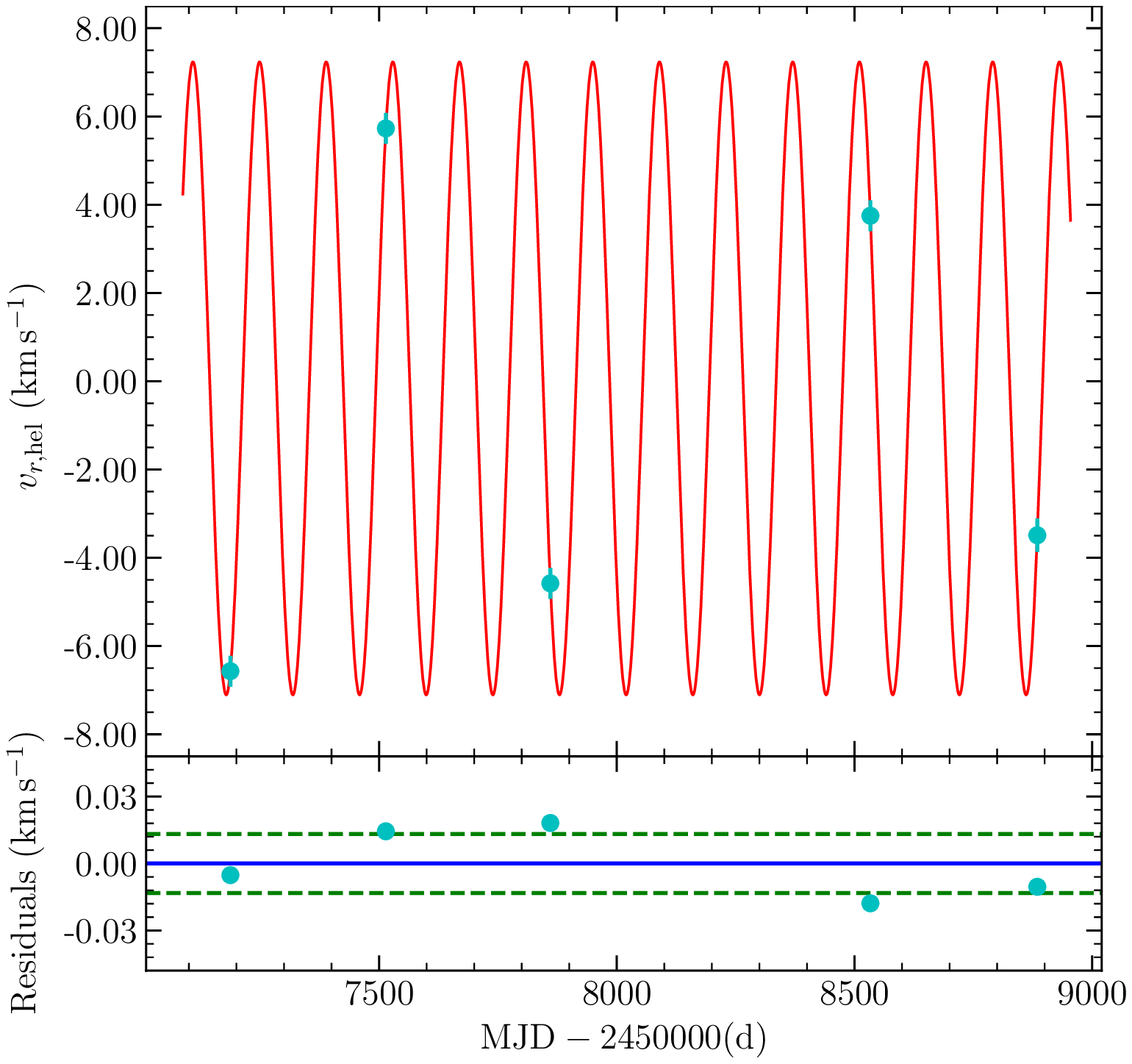}
}
 \caption{Changes in the heliocentric radial velocity of the [N\,{\sc ii}] $\lambda$5755 line in 
 the spectrum of Wray\,15-906. Note that the uncertainties in the data points are very small.}
 \label{fig:vel}
\end{figure}

We also measured the heliocentric radial velocity, $v_{r,{\rm hel}}$, of the [N\,{\sc ii}] $\lambda$5755
line using all five HRS spectra. It is believed that this velocity corresponds to the systemic velocity 
of the star (Stahl et al. 2001). The measurements (made using the programs described 
in Kniazev et al. 2004) are listed in Table\,\ref{tab:rad}. It appears that $v_{r,{\rm hel}}$
changes slightly around zero. We fit these measurements to a sine curve using the $\chi^2$ algorithm. 
The best-fitting result\footnote{We caution that this result is not unique due to the
limited number of data points.} ($\chi^2=7\times10^{-4}$) is shown in Fig.\,\ref{fig:vel}. 
It indicates that $v_{r,{\rm hel}}$ changes
with a period of $140.21\pm0.01$\,d and an amplitude of $7.17\pm0.01 \, \kms$, and implies a
systemic radial velocity of $-0.07\pm0.01 \, \kms$. We speculate that these changes could be 
due to pulsations in the stellar atmosphere (cf. Section\,\ref{sec:evo}).

\begin{table}
\centering\caption{Changes in the heliocentric radial velocity of the [N\,{\sc ii}] $\lambda$5755 
line in the spectrum of Wray\,15-906.}
\label{tab:rad}
\begin{tabular}{lcc} \hline
Date & JD & $V_{\rm r,hel} (\kms)$ \\
\hline
2015 June 13     & 2457187.25163 & $-6.57\pm0.35$ \\
2016 May 5       & 2457514.29555 & $5.73\pm0.35$ \\
2017 April 16    & 2457860.25479 & $-4.58\pm0.35$ \\
2019 February 18 & 2458533.48539 & $3.75\pm0.35$ \\
2020 February 4  & 2458884.49809 & $-3.49\pm0.38$ \\
\hline
\end{tabular}
\end{table}

\section{Spectroscopic and photometric variability of Wray\,15-906}
\label{sec:var}

\subsection{Spectroscopic variability}
\label{sec:spec}

\begin{figure*}
{\centering \resizebox*{0.66\columnwidth}{!}{\includegraphics[angle=0]{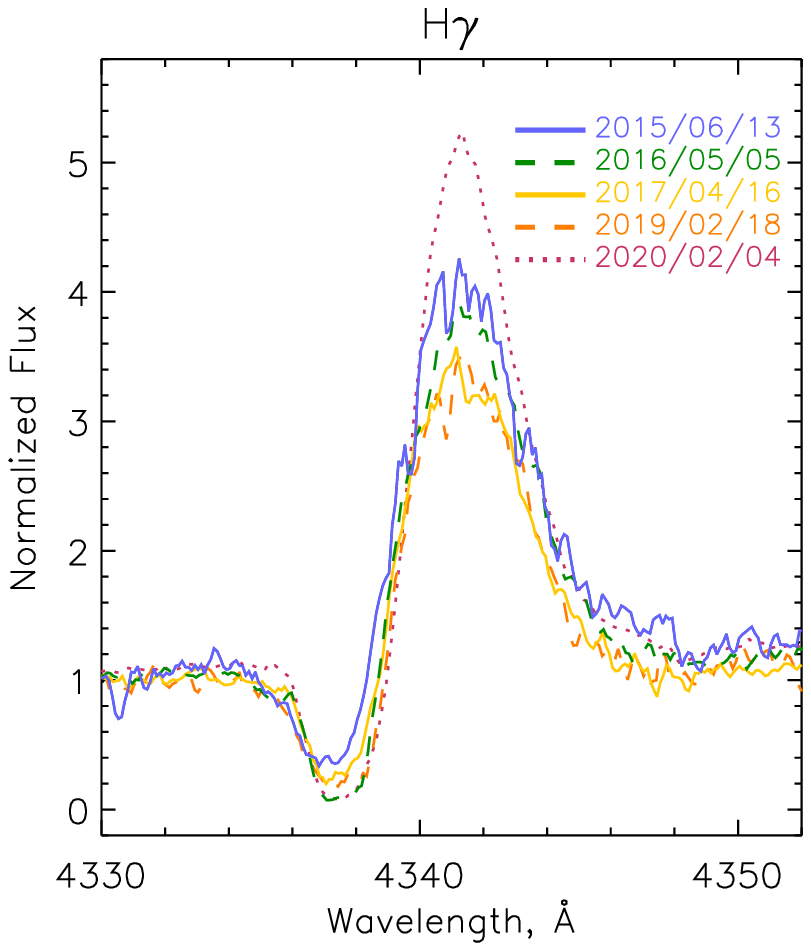}}}
{\centering \resizebox*{0.66\columnwidth}{!}{\includegraphics[angle=0]{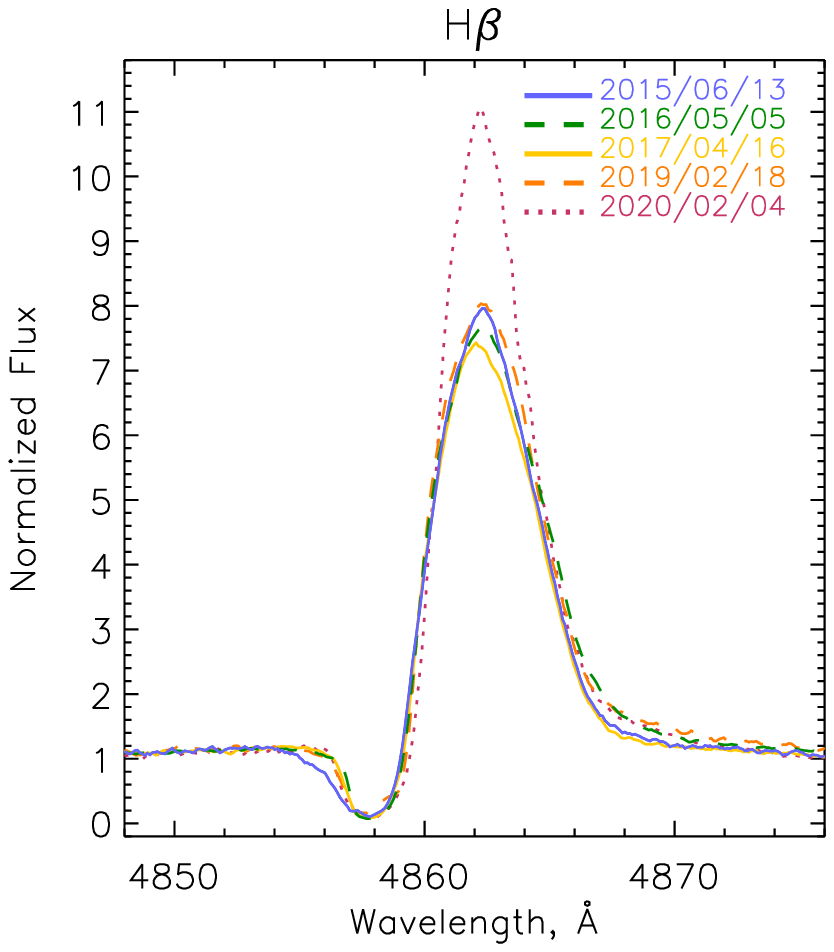}}}
{\centering \resizebox*{0.66\columnwidth}{!}{\includegraphics[angle=0]{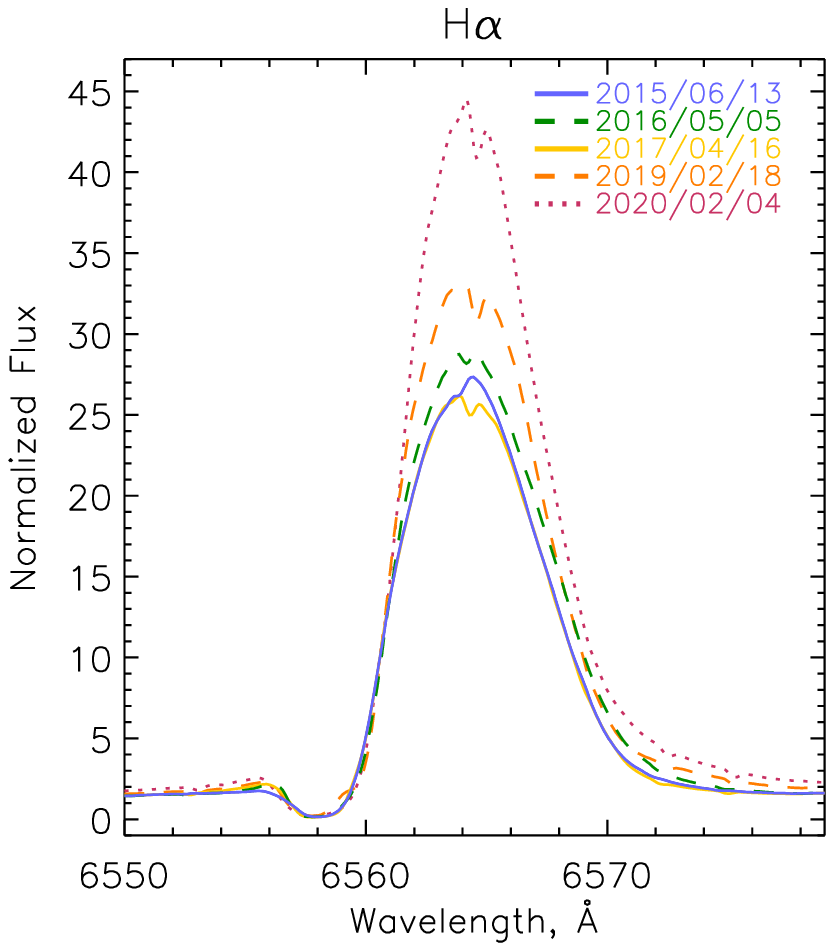}}}\\
{\centering \resizebox*{0.66\columnwidth}{!}{\includegraphics[angle=0]{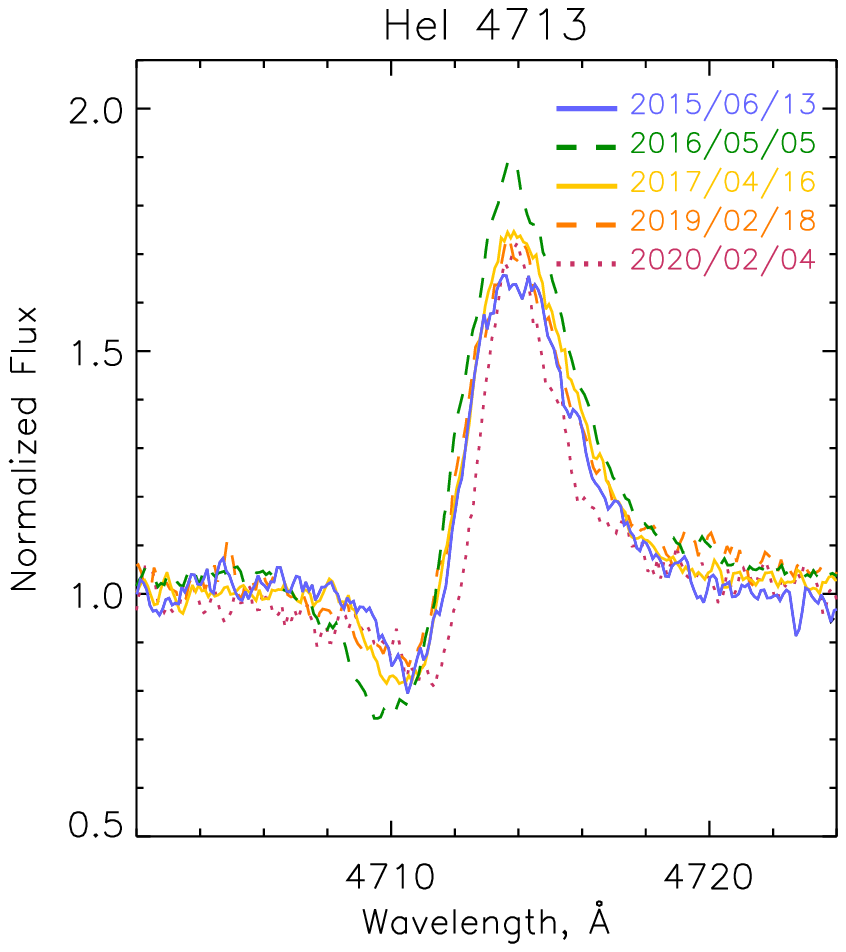}}}
{\centering \resizebox*{0.66\columnwidth}{!}{\includegraphics[angle=0]{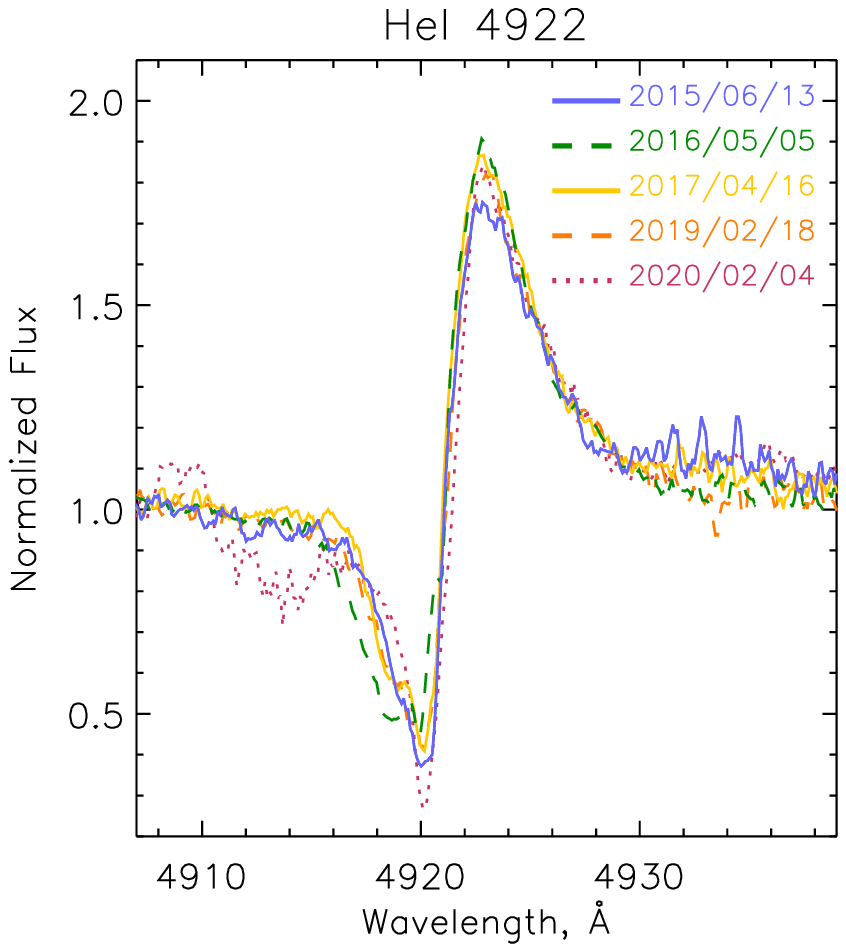}}}
{\centering \resizebox*{0.66\columnwidth}{!}{\includegraphics[angle=0]{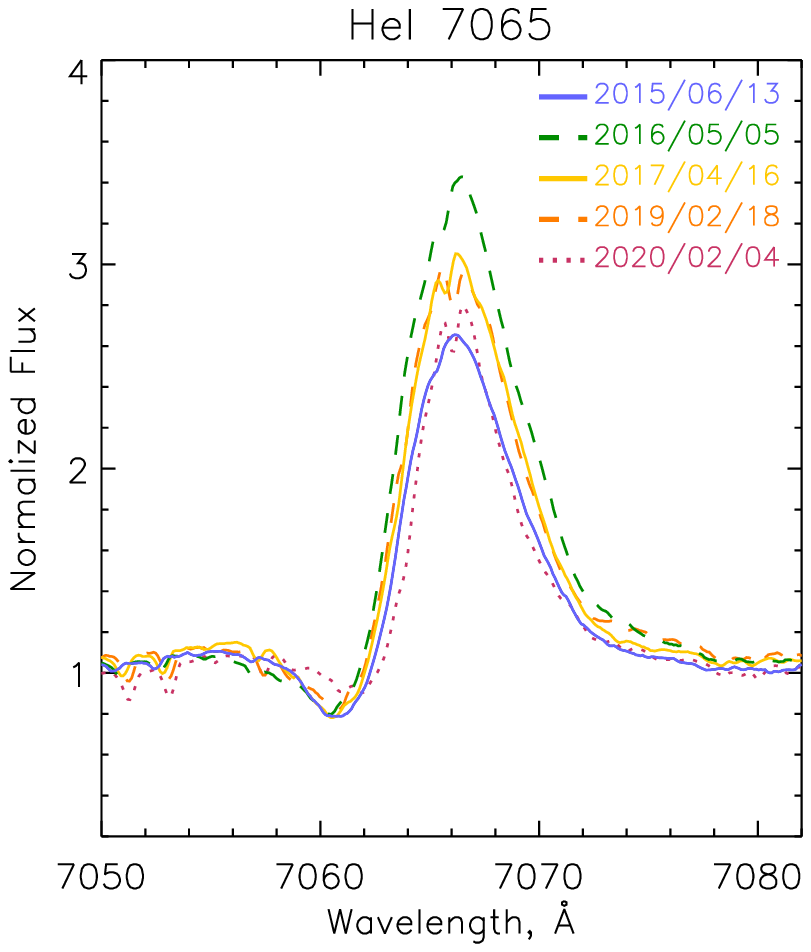}}}\\ 
{\centering \resizebox*{0.66\columnwidth}{!}{\includegraphics[angle=0]{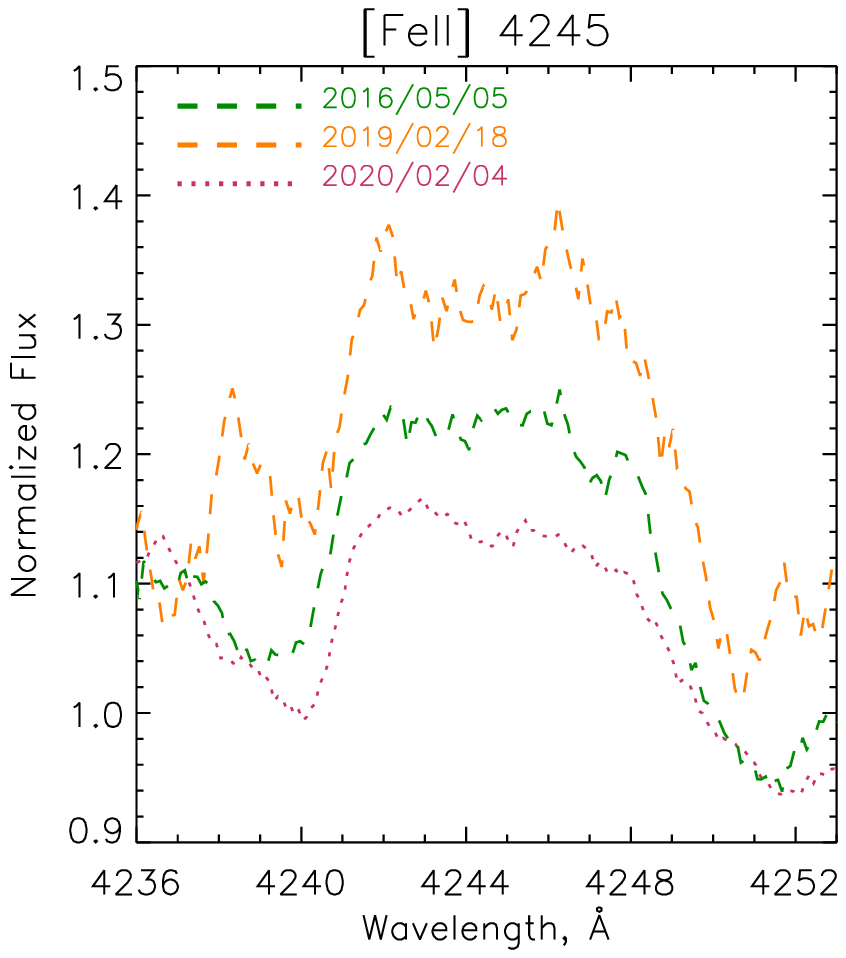}}}
{\centering \resizebox*{0.66\columnwidth}{!}{\includegraphics[angle=0]{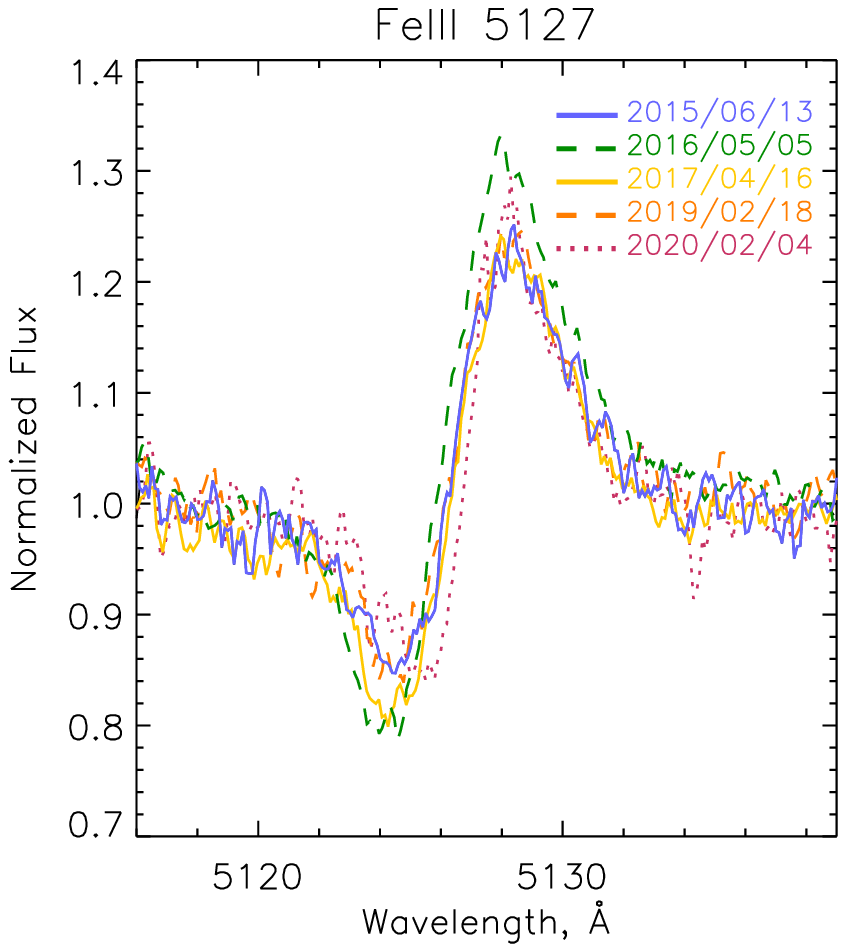}}}
{\centering \resizebox*{0.66\columnwidth}{!}{\includegraphics[angle=0]{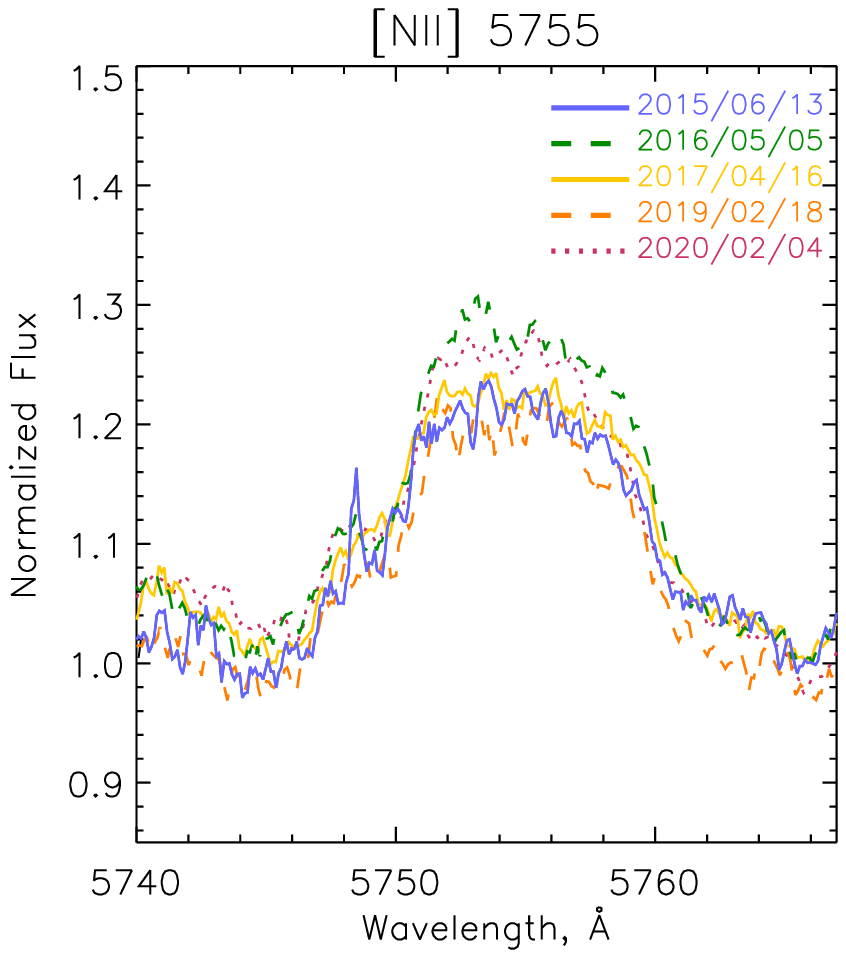}}}\\
\caption{Line profiles variability in the HRS spectra of Wray\,15-906. The dates of observations 
are indicated in the legends.}
\label{fig:var}
\end{figure*}

\begin{figure*}
\begin{center}
\includegraphics[width=8.5cm,angle=0]{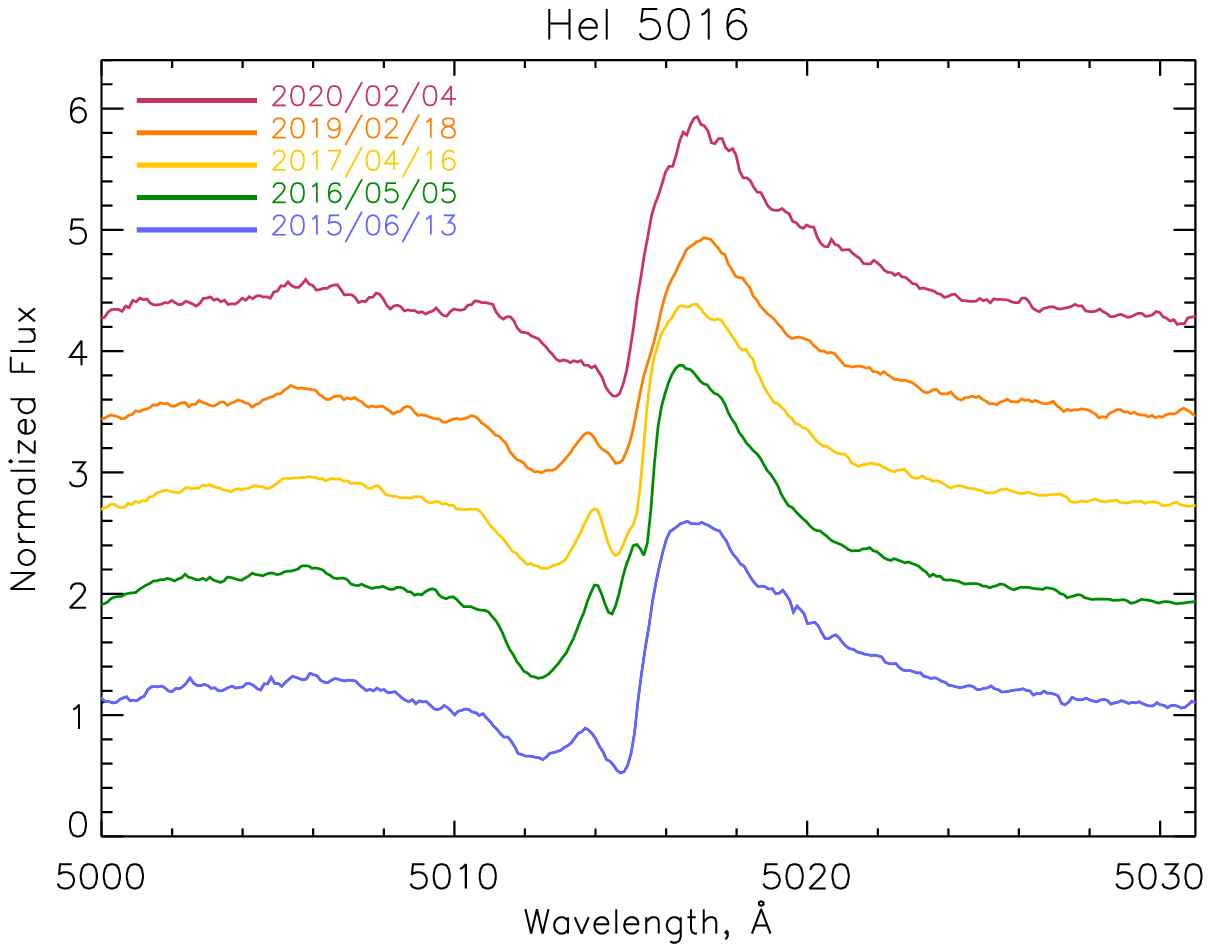}
\includegraphics[width=8.5cm,angle=0]{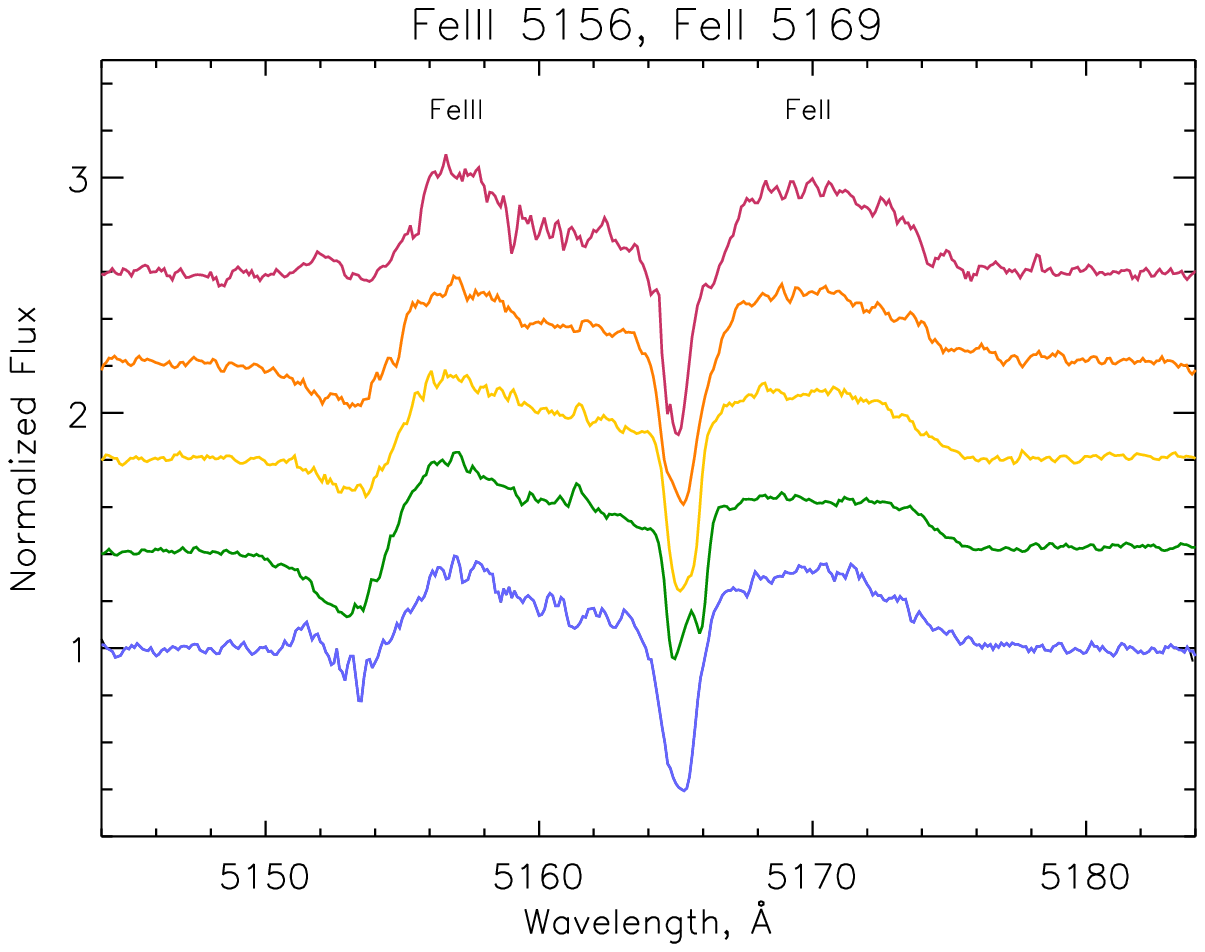}
\end{center}
\caption{Changes in the absorption components of the He\,{\sc i} $\lambda$5016 and 
Fe\,{\sc ii} $\lambda$5169 lines in the HRS spectra taken in 2015--2020.}
\label{fig:abs}
\end{figure*}

As noted in Section\,\ref{sec:neb}, our initial spectroscopic monitoring did not detect significant 
changes in the spectrum of Wray\,15-906 in 2012--2016. Moreover, the spectrum did not change much in 
the following 4 yr. Particularly, the He\,{\sc ii} and N\,{\sc iii} lines did not appear in none of 
the RSS and HRS spectra, meaning that the stellar effective temperature remains nearly constant 
during the 8 yr of observations. Similarly, inspection of the HRS spectra did not reveal noticeable 
changes in the width of flat-topped forbidden lines (see the bottom row in Fig.\,\ref{fig:var}). Since 
the width of these lines is a good measure of the terminal wind velocity (cf. Stahl et al. 1991, 2001), 
we conclude that $v_\infty$ did not change as well. From the radial velocities of the blue and red edges 
of the flat-topped lines, we estimated the wind expansion velocity to be $275\pm30 \, \kms$, which agrees 
well with the wind velocity of $280\pm50 \, \kms$ derived in Section\,\ref{sec:mod}.

On the other hand, the comparison of the spectra revealed moderate variability in the line 
intensities and P\,Cygni profiles, which is more obvious in the HRS spectra. This variability is 
demonstrated in Fig.\,\ref{fig:var} where we plot profiles of selected lines from five HRS spectra 
obtained during the last 4.5 yr. Fig.\,\ref{fig:var} shows that lines experience two types of variability, 
both of which are typical of hot stars with extended atmospheres (Lamers, Korevaar \& Cassatella 1985; 
Kaufer et al. 1996; Markova \& Valchev 2000; Stahl et al. 2001; Chentsov \& Maryeva 2016). 

First of them is the variability of line intensities, more prominent in the Balmer lines (upper row 
in Fig.\,\ref{fig:var}). Particularly, one can see that the intensity of the H\,$\alpha$ line was 
weakly variable in 2015--2018 and increased by $\approx50$ per cent during the last year. This type 
of variability could be attributed to changes in the mass-loss rate, possibly caused by the bi-stability 
jump (Pauldrach \& Puls 1990) or/and stellar pulsations (see Section\,\ref{sec:evo}).

The second type of variability is characterized by the appearance of additional components in the
P\,Cygni absorption profile of some lines, e.g. He\,{\sc i} $\lambda$5016 and Fe\,{\sc ii} $\lambda$5169 
(see Fig.\,\ref{fig:abs}). It could be connected with variability of 
the line-of-sight optical depth of the stellar wind caused by the presence of a large scale spiral-shaped 
pattern within the wind. Such a pattern could arise from interaction of flows of different speed in 
non-spherically symmetric stellar winds (e.g. Kaper et al. 1996; Fullerton et al. 1997).
Appearance of additional P\,Cygni absorption components could also be connected to abrupt changes 
in the wind structure caused by the bi-stability jump (Groh \& Vink 2011), while in binary systems 
they could originate in the wind-wind interaction zone (Lobel et al. 2015).

\subsection{Photometric variability}
\label{sec:phot}

\begin{figure*}
\centering{
 \includegraphics[clip=,angle=-90,width=12cm]{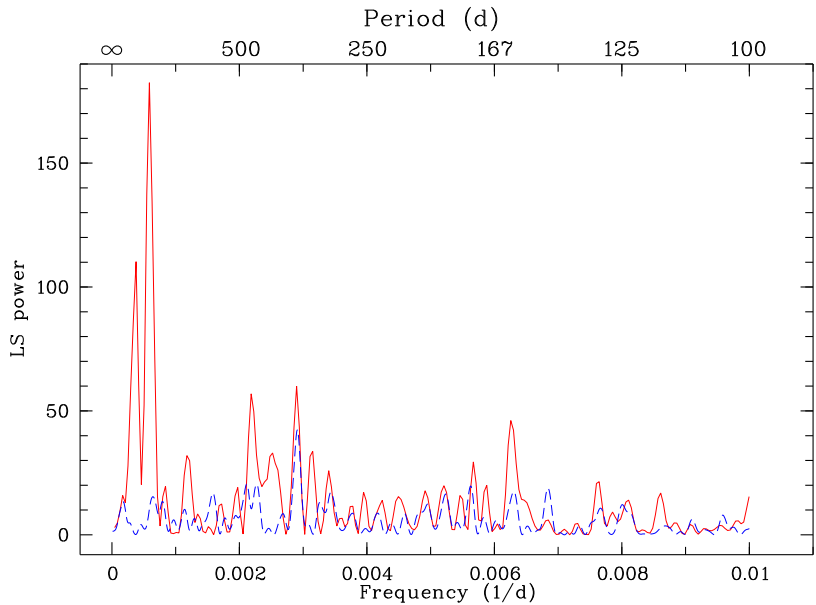}
}
 \caption{A part of the Lomb-Scargle (LS) periodogram of the light curve of Wray\,15-906 from 1999--2019
     showing the strongest power spikes (red solid line). The dashed blue line corresponds to the LS 
     periodogram after subtraction of the three strongest spikes at periods of 1698, 2818 and 395 d. 
     }
 \label{fig:ls}
\end{figure*}

\begin{figure*}
\centering{
 \includegraphics[clip=,angle=-90,width=12cm]{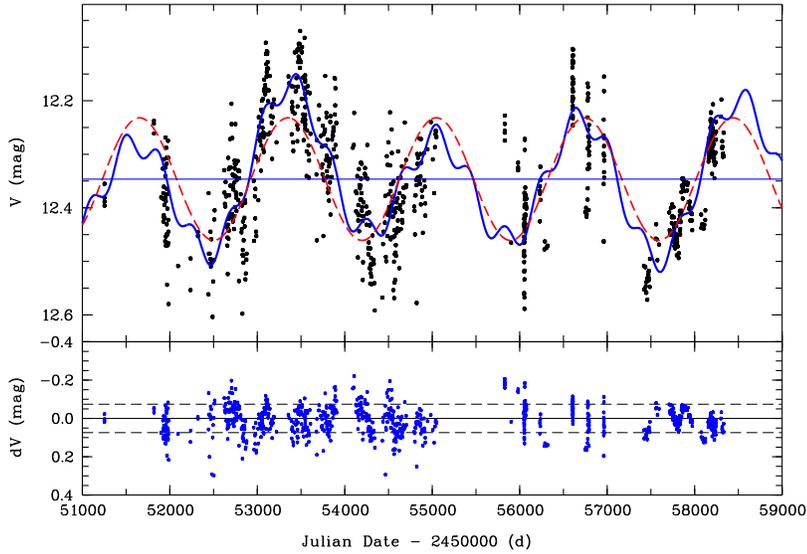}
}
 \caption{Upper panel: the fit to the observed light curve of Wray\,15-906 with three sine 
 waves of period 2818, 1698 and 395 d. The observed light curve is shown with the black dots. 
 The fit to the data is shown with the solid (blue) curve. The dashed (red) sine wave 
 corresponds to the strongest power spike in the LS periodogram at the period of 1698\,d. The 
 horizontal (blue) line  shows the average $V$ magnitude of $12.35\pm0.11$. Bottom panel:  
 residuals of the fit. The mean residual magnitude and its 1$\sigma$ uncertainty of 0.07~mag
 are shown with the solid and dashed lines, respectively.
\label{fig:per}}
\end{figure*}

\begin{figure}
\centering{
 \includegraphics[clip=,angle=-90,width=8.5cm]{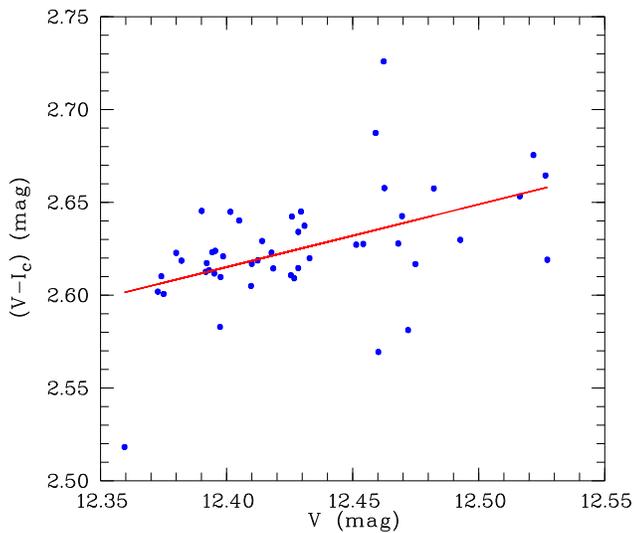}
}
 \caption{$V-I_{\rm c}$ colour of Wray\,15-906 as a function of the $V$ magnitude.
 The solid (red) line is the least squares linear fit.}
\label{fig:col}
\end{figure}
%
\begin{figure}
\centering{
 \includegraphics[clip=,angle=-90,width=8.5cm]{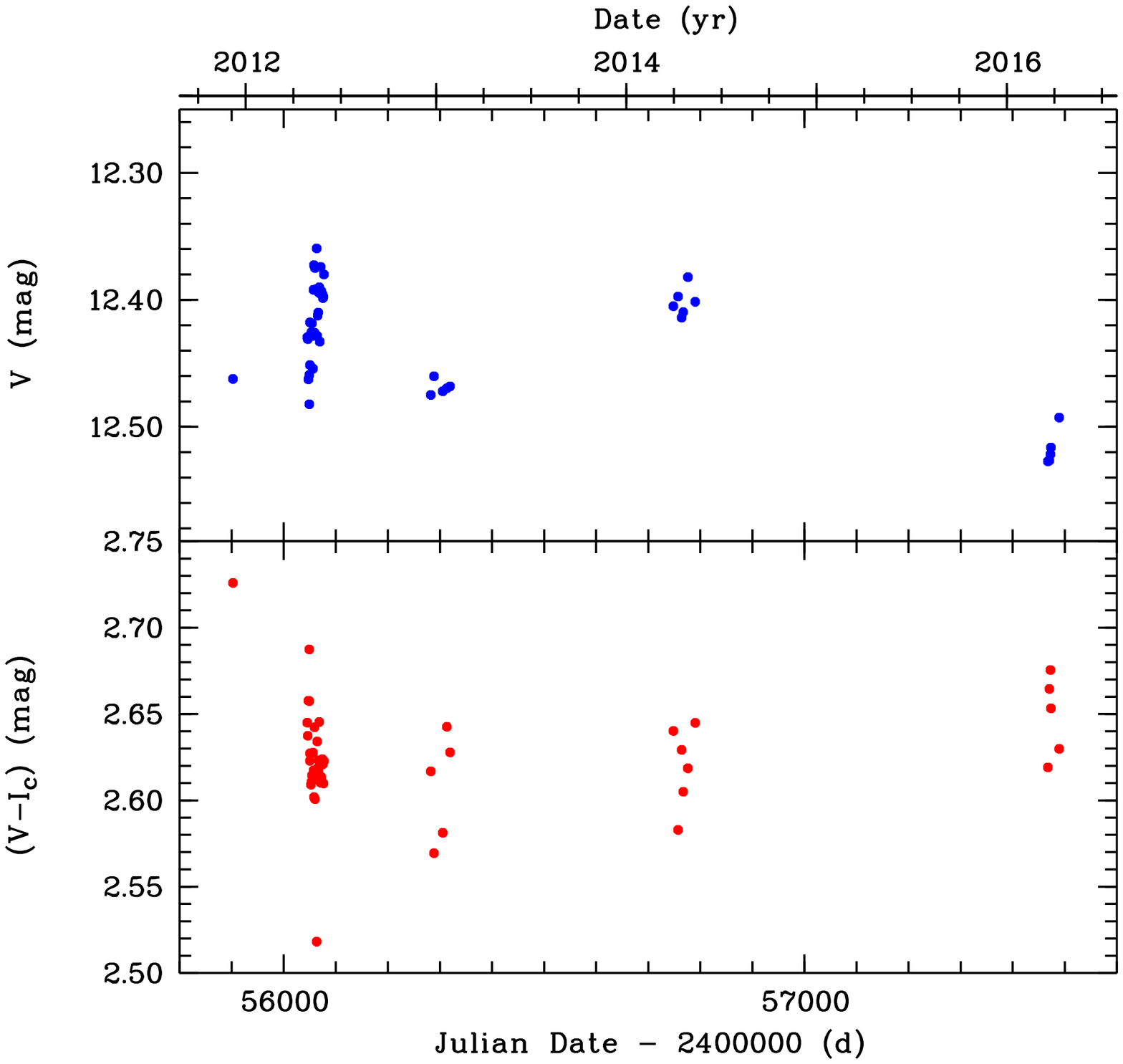}
}
 \caption{Upper panel: Light curve of Wray\,15-906 in the $V$ band in 2012--2016. Bottom panel:  
     Evolution of the $V-I_{\rm c}$ colour of Wray\,15-906 with time.}
\label{fig:colo}
\end{figure}

To search for a possible periodicity in the light curve of Wray\,15-906, we computed its 
Lomb-Scargle (LS) periodogram (Lomb 1976; Scargle 1982) using the {\sc midas} context 
{\sc tsa}. The resulting LS periodogram shows numerous spikes, of which 
the strongest ones are located in the region with periods $>100$\,d. This part of the 
periodogram is shown in Fig.\,\ref{fig:ls}. The most prominent spike corresponds to the 
period of $P_1 = 1698\pm9$\,d. In the upper panel of Fig.\,\ref{fig:per}, we superimposed a sine 
wave with this period (red dashed line) on the observed light curve (black dotes) to show that 
they agree quite well with each other. 

After subtraction of this period from the LS periodogram, we found the next strongest
spike at $P_2=2818\pm110$\,d. Repeating the subtraction procedure leads to the detection of the
third period of $P_3 = 395.31\pm0.22$~d. The LS periodogram after subtraction of the three 
periods is shown in the inset to Fig.\,\ref{fig:ls} with the dashed (blue) line. 

Using the detected three periods, we fit the observed light curve with a linear combination 
of three sine waves with amplitudes $A_1=0.112\pm0.004$\,mag, $A_2=0.038\pm0.004$\,mag and 
$A_3=0.025\pm0.003$\,mag The resulting synthetic light curve is shown with a solid blue line 
in the upper panel of Fig.\,\ref{fig:per}. The bottom panel of Fig.\,\ref{fig:per} shows 
residuals of the fit (the dispersion of the residuals is 0.07~mag).

Then, we re-analysed the light curve after excluding data points with the lowest accuracy
(i.e. the OMC photometry). In this case, the LS periodogram shows the strongest power spike at 
a period of $1729\pm3$\,d, which is almost identical to the period $P_1$, while the 
two other strongest spikes were found at periods of $7422\pm20$ and $289\pm4$\,d.
This exercise confirms that the period of $\approx1700$\,d is indeed present in the light curve,
and suggests that the other two periods are rather spurious. Although it is not clear what the
origin of the $\approx1700$\,d period, it is tempting to assume that it is due to possible duplicity 
of the star, i.e. related to the orbital period of the binary system. Further coordinated photometric
and spectroscopic observations of Wray\,15-906 along with long baseline interferometric imaging
are needed to confirm or reject this assumption (cf. Richardson, Gies \& Williams 2011; Boffin et al., 
2016).

Also, we used our photometric observations in 2011--2016 to examine whether Wray\,15-906
has experienced LBV excursions in the Hertzsprung-Russell diagram during this time period, in which 
case the star is expected to become bluer (hotter) with the brightness decrease and redder (cooler) 
when its brightness increases (e.g. van Genderen 1982; see also figs 5 and 6 in Kniazev et al. 2016). 
Fig.\,\ref{fig:col} plots the $V-I_{\rm c}$ colour of Wray\,15-906 as function of the $V$ magnitude. 
What we see is that the star becomes redder with the brightness decrease, which is opposite to what 
one expects for the LBV excursions. This effect is also evident in Fig.\,\ref{fig:colo} showing changes 
in the $V$ magnitude and the $V-I_{\rm c}$ colour with time. Since we did not detect noticeable changes 
in the spectral appearance of the star (meaning that $T_{\rm eff}$ was almost constant during our 
observations), we attribute the changes in the colour to variability of circumstellar extinction, 
probably caused by variability of the mass-loss rate.

\section{Discussion}
\label{sec:dis}

\subsection{Evolutionary status and possible fate of Wray\,15-906}
\label{sec:evo}

\begin{figure}
\begin{center}
\includegraphics[width=8.5cm,angle=0]{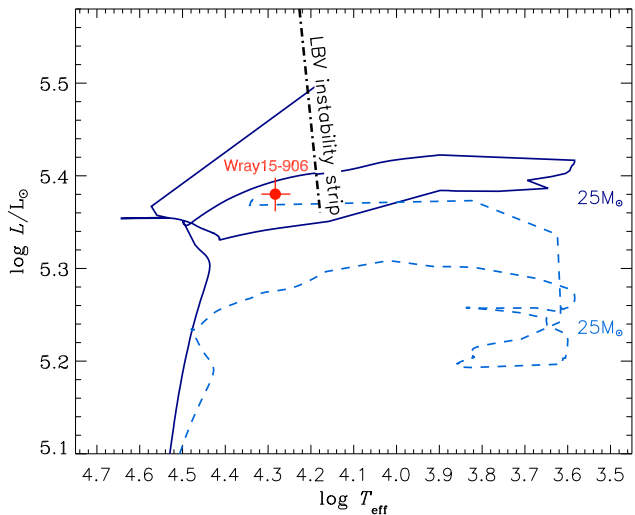}
\end{center}
\caption{Position of Wray\,15-906 in the Hertzsprung-Russell diagram (red dot with error bars). 
The dashed and solid lines show the Geneva evolutionary tracks of a $25 \, \msun$ single, solar-metallicity 
star with initial rotational velocities of 0 and 40 per cent of breakup, respectively (Ekstr\"om et 
al. 2012). The dot-dashed line marks the location of the LBV instability strip as defined in Groh 
et al. (2009a).}
\label{fig:hrd}
\end{figure}

Fig.\,\ref{fig:hrd} shows the position of Wray\,15-906 in the Hertzsprung-Russell diagram for the
adopted distance of 3.53 kpc along with the Geneva evolutionary tracks of single, solar-metallicity 
non-rotating and rotating (at 40 per cent of breakup) stars with initial mass of $M_{\rm init}=25 
\, \msun$ (Ekstr\"om et al. 2012). If Wray\,15-906 was born single, then its initial mass could be 
estimated to be $25 \, \msun$, which is close to the minimum mass for single (non-rotating) 
stars to become a Wolf-Rayet star (e.g. Georgy et al. 2012). According to the Geneva models, a 
star of this $M_{\rm init}$ loses about 70 per cent of its mass during the red supergiant phase 
and then evolves blueward in the Hertzsprung-Russell diagram to become a Wolf-Rayet star. 

The surface He abundance (65 per cent by mass) derived for Wray\,15-906 
is typical of LBVs (e.g. Crowther 1997) and is intermediate between that of O and Wolf-Rayet stars. This
suggests that Wray\,15-906 has already gone through the red supergiant phase and currently is 
near the end of the He-burning stage, and implies the evolutionary age of this star 
of $\approx7-8$ Myr. As a blue supergiant evolving from the the red supergiant stage, 
Wray\,15-906 should be subject to radial pulsations (e.g. Saio, Georgy \& Meynet 2013; 
Jeffery \& Saio 2016), which could be responsible for the heliocentric radial velocity 
variability in this star (see Table\,\ref{tab:rad}) and for its brightness variations on 
time-scales of the order of 100\,d.

If Wray\,15-906 was born with a low rotational velocity then it would need
only several 1000\,yr to finish its life with a supernova explosion (Georgy et 
al. 2012). The possibility that Wray\,15-906 may soon explode is indirectly supported by the
curious finding by Groh, Meynet \& Ekstr\"om (2013a) and Groh et al. (2013b) that single stars 
with $M_{\rm init}=20-25 \, \msun$ have LBV-like   
spectra before exploding as a supernova. Although unexpected, this finding conforms with previous
claims (e.g. Kotak \& Vink 2006) that LBVs could be the immediate precursors of supernovae. 

Groh et al. (2013a,b) used the stellar evolution models by Ekstr\"om et al. (2012) as input in
{\sc cmfgen} stellar atmosphere models to compute synthetic optical spectra and photometry (absolute
magnitudes, colours and bolometric corrections) in different filters at the pre-supernova stage for 
stars with $M_{\rm init}=9-120 \, \msun$. They found that 
pre-supernova spectra of rotating models with $M_{\rm init}=20-25 \, \msun$ are very similar to 
those of bona-fide LBVs, such as P\,Cygni and AG\,Car. They also found that their non-rotating model
with $M_{\rm init}=25 \, \msun$ appears at death as a WN11h star, such as Wray 15-682 or AG\,Car at
its visual minimum.
In this connection, we recall (see Section\,\ref{sec:ste} and Fig.\,\ref{fig:com}) that the spectrum of 
Wray\,15-906 bears a strong resemblance with that of Wray 15-682, which given the blueward evolution of
Wray\,15-906 suggests that this star may end its life as a WN11h star.

\begin{table}
\caption{Comparison of Wray\,15-906 with non-rotating (`nr') and rotating (`r') $25 \, \msun$ 
model stars at the pre-supernova stage from Ekstr\"om et al. (2012) and Groh et al. (2013b).}
\label{tab:com}
\begin{center}
\begin{tabular}{lccc}
\hline 
                              & Wray\,15-906 & nr & r \\
\hline
$\log(L_*/\lsun)$             & 5.36--5.40  & 5.38    & 5.50 \\
$T_*$ (kK)                    & 23--27      & 27.1    & 24.6 \\
$T_{\rm eff}$ (kK)            & 17.7--20.7  & 26.3    & 20.9 \\
He (mass fraction)            & 0.63--0.67  & 0.83    & 0.93 \\
C (mass fraction) (10$^{-5}$) & $\leq6$     & 6.8     & 16   \\
N (mass fraction) (10$^{-3}$) & 2.7--3.5    & 8.2     & 16   \\ 
O (mass fraction) (10$^{-4}$) & 0.8--1.6    & 1.1     & 6.5  \\
\hline
\end{tabular}
\end{center}
\end{table}

In Table\,\ref{tab:com}, we compare some parameters of Wray\,15-906 with corresponding parameters of 
non-rotating and rotating (at 40 per cent of breakup) $25 \, \msun$ models at the pre-supernova stage 
compiled from Ekstr\"om et al. (2012) and Groh et al. (2013b)\footnote{Note a misprint in table\,3 of 
this paper, where the N abundance in the rotating $25 \, \msun$ model should have a tenfold higher value.}.
One can see that  most parameters of Wray\,15-906 agree fairly well with parameters of the non-rotating
model with the main difference being that the model star appears hotter and its He abundance is much
higher, which is however expectable because after the post-He-burning stage the temperature and the 
mass fraction of He at the surface of non-rotating stars increase to their pre-supernova values on 
a time scale of $\sim1000$\,yr. Table\,\ref{tab:com} also shows that the N abundance derived for 
Wray\,15-906 is about a factor of two lower than what follows from the non-rotating model. Although 
this discrepancy, at least in part, may be due to a flaw in our spectral modelling, it should be 
noted that abundances predicted by stellar evolution models are very sensitive to the adopted 
prescriptions for convection, rotational mixing and mass-loss rates, which are still not well known
and understood. We therefore speculate that Wray\,15-906 may soon turn into a WN11h
star before core collapse. If our speculation is correct, then one can expect that Wray\,15-906
will explode as a supernova before its circumstellar shell dissipates into the interstellar medium.
In this case, after $\sim100$\,yr the supernova blast wave will catch up and interact with the 
shell to produce a pc-scale supernova remnant. 

It should be noted, however, that recent studies of red supergiants in star clusters (Beasor et al. 
2020 and references therein) suggest that their mass-loss rates are more than an order of magnitude 
lower than those adopted in Ekstr\"om et al. (2012) and Groh et al. (2013a,b). If correct, this 
result would imply that during the red supergiant phase single stars did not lose enough mass to 
evolve blueward. However, it did not exclude the possibility that these stars can experience episodes 
of extreme mass-loss at the end of the red supergiant phase (e.g. Yoon \& Cantiello 2010) or during
the subsequent yellow hypergiant phase in the course of excursions on the cool side of the yellow 
void (de Jager \& Nieuwenhuijzen 1997).
Moreover, the mass loss required for the blueward evolution could also be driven by a companion star 
(e.g. through common-envelope ejection), which is quite possible given the high proportion of binaries 
among massive stars (Sana et al. 2012). The presence of the circumstellar shell around Wray\,15-906 
indicates that this star may have suffered mass loss at high rate in the recent past. To prove this, let 
us estimate the mass of this shell.

\subsection{Mass of the circumstellar shell}
\label{sec:mass}

To estimate the mass of the shell, we measured its flux densities at 22 and 70\,\micron \,
using an 135 arcsec aperture centred on Wray\,15-906. After subtraction contributions from the star
and background emission, we found $F_\nu=0.80\pm0.01$\,Jy at 22\,\micron \, and $F_\nu=9.04\pm0.94$\,Jy 
at 70\,\micron. With these fluxes in hands, one can estimate the mass, $M_{\rm d}$, and temperature,
$T_{\rm d}$, of the dust in the shell using the following relationship (Hildebrand 1983):
\begin{equation}
M_{\rm d}={F_\nu d^2\over B(\nu,T_{\rm d})\kappa_\nu} \, ,
\label{eq:flux}
\end{equation}
where $B(\nu,T_{\rm d})$ is the Planck function and $\kappa_\nu$ is the mass absorption coefficient 
at frequency $\nu$, which for $\lambda >10$\,\micron \, is given by (Laor \& Draine 1993; Meikle et 
al. 2007):
\begin{equation}
\kappa_\nu\approx1000 \left({\lambda\over 20\,\micron}\right)^{-2} \, {\rm cm}^2 \, {\rm g}^{-1} \, .
\label{eq:kappa}
\end{equation}
Using equations (\ref{eq:flux}) and (\ref{eq:kappa}) for the wavelengths 22 and 70\,micron,  
one finds $T_{\rm d}\approx55\pm1$\,K, while adopting $d\approx3.53$ kpc results in  
$M_{\rm d}\approx(1.45\pm0.25)\times 10^{-2} \, \msun$. The later estimate translates into the 
mass of the shell of $M_{\rm sh}\approx2.9\pm0.5 \, \msun$ 
if one assumes the gas-to-dust mass ratio of 200 (typical of Galactic red supergiants; 
Mauron \& Josselin 2011). Interestingly, the mass and radius ($\approx1$\,pc)
of the shell around Wray\,15-906 are similar to those of the shell around Wray\,15-682 of, respectively,
$2\pm1 \, \msun$ and $\approx1$\,pc (Smith et al. 1994). This further strengthens the similarity 
between the two stars and makes it reasonable to assume that the kinematic ages of both shells
are similar as well ($\sim10,000$\,yr; Smith et al. 1994). 

\begin{table*}
\caption{Kinematic data on Wray\,15-906 (see text for details).}
\label{tab:pro}
\begin{tabular}{cccccc}
\hline 
$d$   & $v_{\rm l}$   & $v_{\rm b}$    & $v_{\rm r}$    & $v_{\rm tr}$  & $v_*$    \\
(kpc) & ($\kms$)      & ($\kms$)       & ($\kms$)       & ($\kms$)      & ($\kms$) \\
\hline 
3.00  & $19.0\pm1.3$ & $5.5\pm1.0$  & 18.5 & $19.8\pm1.3$  & 27.1 \\ 
3.53  & $21.8\pm1.5$ & $5.2\pm1.2$  & 19.3 & $24.4\pm1.4$  & 31.1 \\ 
4.27  & $25.7\pm1.8$ & $4.6\pm1.5$  & 18.2 & $26.1\pm1.8$  & 31.8 \\ 
\hline
\end{tabular}
\end{table*}

\begin{figure*}
\begin{center}
\includegraphics[width=15cm,angle=0]{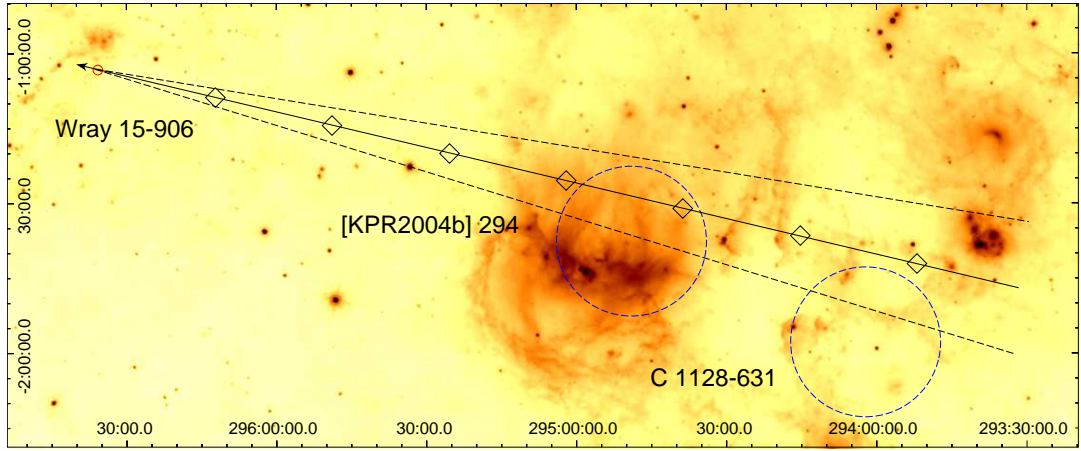}
\end{center}
\caption{{\it WISE} 22\,\micron \, image of the field containing Wray\,15-906 (marked by a red circle) 
and two open star clusters [KPR2004b] 294 and C 1128-631 (indicated by dashed blue circles). The arrow 
shows the direction of motion of Wray\,15-906 for $d=3.53$\,kpc. A solid line shows the trajectory of 
Wray\,15-906 with $1\sigma$ uncertainties shown by dashed lines. Diamonds mark the positions of 
Wray\,15-906 over a time period of 7\,Myr with a time step of 1\,Myr. The image is oriented with Galactic 
longitude (in units of degrees) increasing to the left and Galactic latitude increasing upwards. 
At the distance of 3.53 kpc, $1\degr$ corresponds to $\approx60.7$ pc.}
\label{fig:clu}
\end{figure*}

\subsection{Wray\,15-906 as a runaway}
\label{sec:run}

Like the majority of (c)LBVs, Wray\,15-906 is located outside of known star clusters
and therefore is likely a runaway star (Gvaramadze et al. 2012b). Since {\it Gaia} DR2
provides accurate proper motion measurements for Wray\,15-906 (see Table\,\ref{tab:det}), it 
is tempting to calculate its past trajectory and to search for its possible parent star cluster.
For this we used the solar peculiar motion of $(U_{\odot},V_{\odot},W_{\odot})=(11.1,12.2,7.3) 
\, \kms$ (Sch\"onrich, Binney \& Dehnen 2010) and the solar Galactocentric distance of $R_0 = 
8.0$ kpc and the circular Galactic rotation velocity of $\Theta _0 =240 \, \kms$ (Reid et al. 2009).
Table\,\ref{tab:pro} gives the peculiar velocity components along the Galactic longitude ($v_{\rm l}$) 
and latitude ($v_{\rm b}$), and the peculiar transverse velocity of Wray\,15-906 of 
$v_{\rm tr}=(v_{\rm l}^2 +v_{\rm b}^2)^{1/2}$ for three values of the distance. To this table
we also added the peculiar radial velocity ($v_{\rm r}$) of Wray\,15-906, derived under the assumption 
that the heliocentric radial velocity of this star is equal to zero (see Section\,\ref{sec:mod}), and the 
total space velocity ($v_*$) of the star. For the error calculation, only uncertainties in the proper 
motion measurements were taken into account. Note that the total space velocity of Wray\,15-906
only weakly depends on the distance. One can show that for $d$ ranging from, say, 2 to 6 kpc, $v_*$ 
ranges from $\approx20$ to $30 \, \kms$.

Table\,\ref{tab:pro} shows that Wray\,15-906 is moving almost parallel to the Galactic plane
with a velocity of $\approx20-25 \, \kms$. The total space velocity of Wray\,15-906 of $\approx30 
\, \kms$ is typical of classical runaway stars (Blaauw 1961). In Fig.\,\ref{fig:clu}, we plot the 
past trajectory of Wray\,15-906 over a time period of 7\,Myr. Inspection of the  
WEBDA\footnote{http://webda.physics.muni.cz/} data base (Mermilliod 1995) showed that there are 
two young star clusters, [KPR2004b]\,294\footnote{It is believed (Schilbach \& R\"oser 2008) that 
[KPR2004b]\,294 is the birth cluster of the O6.5\,III(f) runaway star HD\,96946.} and C\,1128-631, 
whose approximate boundaries (shown by dashed blue circles of radius of $0\fdg25$; Piskunov et al. 
2007) are crossed by the trajectory of Wray\,15-906. According to WEBDA, [KPR2004b]\,294 and 
C\,1128-631 are located at the distances of 2.74 and 3.40 kpc, and their ages are $\approx4.4$ and 
15 Myr, respectively. Although, due to the zero-point problem in the {\it Gaia} DR2 parallaxes (e.g.
Lindegren et al. 2018), one cannot rule out the possibility that Wray\,15-906 is located at 
the same distance as [KPR2004b]\,294, the much younger age of the latter makes the physical 
relationship between the two objects unlikely. Moreover, at the shorter distance the evolutionary 
age of Wray\,15-906 (implied by the He abundance on the surface of this star) would be higher, 
which makes the age discrepancy even more worse.  

The distance to C\,1128-631 is in a good agreement with that to Wray\,15-906, but the cluster is
a factor of about two older than the star. This implies that either the two objects are not related
to each other or that Wray\,15-906 is a rejuvenated product of binary evolution, i.e. a blue straggler
(e.g. Vanbeveren et al. 2013; Langer \& Kudritzki 2014, Podsiadlowski \& Vink 2014). In this case, 
the star most likely will not evolve through the red supergiant phase and will remain in the blue part 
of the Hertzsprung-Russell diagram. Also, one cannot exclude the possibilities that the parent cluster 
of Wray\,15-906 has already dissolved or that this star has found itself in the field because of the 
two-step-ejection process, i.e. due to the dissolution of a runaway massive binary (Pflamm-Altenburg \& 
Kroupa 2010). In the latter case, the parent cluster cannot be determined at all.

Finally, we note that the central location of Wray\,15-906 within the almost perfectly circular 
shell implies that the shell does not feel the ram pressure of the interstellar medium caused by 
the motion of the star. This could be understood if the nebula was ejected (almost) instantly (e.g. 
due to some catastrophic event in the underlying star or due to a brief episode of high mass loss),
in which case the ejecta should be massive enough not to feel the ambient medium. This implies that 
the mass of the shell should be much larger than the mass of the interstellar medium displaced by the 
shell. Adopting $M_{\rm sh}=2.9 \, \msun$, one finds that the number density of the interstellar 
medium should be $<<20 \, {\rm cm}^{-3}$.
Alternatively, the shell around Wray\,15-906 could be the result of interaction between the current 
stellar wind and a co-moving dense material lost by the star during the preceding red supergiant phase. 
In this case, the wind-wind interaction region (shell) could be shielded from the ram pressure of the 
interstellar medium by the external material of the red supergiant wind (cf. Gvaramadze et al. 2019a).

\section{Summary}
\label{sec:sum}

We have presented the results of study of the new Galactic cLBV Wray\,15-906 discovered via the 
detection of its circular circumstellar shell with {\it WISE} and {\it Herschel}, and follow-up 
spectroscopy with SALT. Our long-slit and \'echelle spectroscopic monitoring of Wray\,15-906 
lasting 8 years (from 2012 February to 2020 February) did not reveal significant changes in the 
spectrum of this star, but resulted in detection of moderate variability in the intensities and 
P\,Cygni absorption profiles of some lines, which is typical of hot stars with extended atmospheres. 
Also, we did not find significant changes in the brightness of Wray\,15-906 over the time
period from 2001 to 2019. The $V$-band light curve constructed using archival and our own photometric 
observations revealed quasi-periodic variability with a period of $\approx1700$\,d and an amplitude 
of $\approx0.1$ mag. We analysed one of the \'echelle spectra of Wray\,15-906 with the stellar 
atmosphere code {\sc cmfgen}, obtaining a stellar temperature of 25\,kK. Wray\,15-906 is highly 
reddened, $E(B-V)\approx2$\,mag. Using the {\it Gaia} DR2 distance to Wray\,15-906 of 3.53 kpc, we 
derived the stellar luminosity of $\log(L/\lsun)\approx5.4$ and the mass-loss rate of 
$\approx3\times10^{-5} \, \myr$. The stellar wind composition is dominated by helium ($\approx65$ per 
cent by mass), implying that Wray\,15-906 is a blueward evolving post-red supergiant star. If Wray\,15-906 
was born single, then its initial mass was $25 \, \msun$, which is close to the minimum mass for 
single (non-rotating) stars to become a Wolf-Rayet star. We have speculated that Wray\,15-906 may soon 
transform into a WN11h star to explode as a supernova after $\sim1000$\,yr. We estimated the dust 
temperature and mass in the circumstellar shell to be $\approx55$\,K and $0.015 \, \msun$, respectively, 
implying the mass of the shell of about $3 \, \msun$ for the gas-to-dust mass ratio of 200. The presence 
of the shell of this mass indicates that Wray 15-906 has experienced a substantial mass loss in the 
recent past, which allowed this star to evolve blueward in the Hertzsprung-Russell diagram. We have 
calculated the peculiar transverse velocity of Wray\,15-906 to be $\approx25 \, \kms$ using {\it Gaia} 
DR2 proper motion measurements and searched for the possible parent star cluster for this star within 
the error cone of its past trajectory. We found that about 7 Myr ago, the trajectory of Wray\,15-906 
passed through the open star cluster C1128-631, whose age is about twice the age of the star, meaning 
that the two object could be associated with each other if Wray\,15-906 is a rejuvenated product of 
binary evolution (a blue straggler).

\section{Acknowledgements}
This work is based on observations obtained with the Southern African Large Telescope (SALT), 
programmes 2013-1-RSA\_OTH-014, 2013-2-RSA\_OTH-003, 2015-1-SCI-017, 2016-1-SCI-012, 2017-1-SCI-006
and 2018-1-MLT-008, and supported by the Russian Foundation for Basic Research grant 19-02-00779. 
O.M. acknowledges support from the Czech Science Foundation GA18-05665S. A.Y.K. acknowledges support 
from the National Research Foundation (NRF) of South Africa. We are grateful to F.\,Najarro for useful 
discussion and for providing us with a {\sc cmfgen} model of P\,Cygni. This research has made use of 
the NASA/IPAC Infrared Science Archive, which is operated by the Jet Propulsion Laboratory, California 
Institute of Technology, under contract with the National Aeronautics and Space Administration, the 
SIMBAD data base and the VizieR catalogue access tool, both operated at CDS, Strasbourg, France, 
the WEBDA data base, operated at the Department of Theoretical Physics and Astrophysics of the Masaryk 
University, and data from the European Space Agency (ESA) mission {\it Gaia} 
(https://www.cosmos.esa.int/gaia), processed by the {\it Gaia} Data Processing and Analysis Consortium 
(DPAC, https://www.cosmos.esa.int/web/gaia/dpac/consortium). Funding for the DPAC has been provided by 
national institutions, in particular the institutions participating in the {\it Gaia} Multilateral 
Agreement.

\section{Data availability}

The data underlying this article will be shared on reasonable request to the corresponding authors.

\end{document}